\documentclass[12pt,preprint]{aastex}

\newcommand{\noprint}[1]{}
\newcommand{\figsetstart}{{\bf Fig. Set} }
\newcommand{\figsetend}{}
\newcommand{\figsetgrpstart}{}
\newcommand{\figsetgrpend}{}
\newcommand{\figsetnum}[1]{{\bf #1.}}
\newcommand{\figsettitle}[1]{ {\bf #1} }
\newcommand{\figsetgrpnum}[1]{\noprint{#1}}
\newcommand{\figsetgrptitle}[1]{\noprint{#1}}
\newcommand{\figsetplot}[1]{\noprint{#1}}
\newcommand{\figsetgrpnote}[1]{\noprint{#1}}

\begin {document}

\title{The Role of Evolutionary Age and Metallicity in the Formation of Classical Be Circumstellar Disks I. 
New Candidate Be Stars in the LMC, SMC, and Milky Way}

\author{J.P. Wisniewski\altaffilmark{1,3} \& K.S. Bjorkman\altaffilmark{2,3}}

\altaffiltext{1}{Universities Space Research Association/NASA GSFC Code 667, Building 21, Greenbelt, MD 20771, jwisnie@milkyway.gsfc.nasa.gov}
\altaffiltext{2}{Ritter Observatory, Department of Physics and Astronomy MS 113, University of Toledo, Toledo, OH 43606, karen@physics.utoledo.edu}
\altaffiltext{3}{Visiting Astronomer, Cerro Tololo Inter-American Observatory}

\begin{abstract}

We present B, V, R, and H$\alpha$ photometry of 8 clusters in the 
Small Magellanic Cloud, 5 in the Large Magellanic Cloud, 
and 3 Galactic clusters, and use  
2 color diagrams (2-CDs) to identify candidate Be star populations in these clusters.  We find 
evidence that the Be 
phenomenon is enhanced in low metallicity environments, based on the observed fractional 
early-type candidate Be star content of clusters of age 10-25 Myr.  Numerous candidate Be stars of
spectral types B0 to B5 were identified in clusters of age 5-8 Myr, challenging the 
suggestion of \citet{fab00} that classical Be stars should only be found in 
clusters at least 10 Myr old.  These results suggest that a significant number of B-type stars must 
emerge onto the zero-age-main-sequence as rapid rotators.  We also detect an enhancement in the
fractional content of early-type candidate Be stars in clusters of age 10-25 Myr, suggesting that
the Be phenomenon does become more prevalent with evolutionary 
age.  We briefly discuss the mechanisms which 
might contribute to such an evolutionary effect.  A discussion of the limitations of 
utilizing the 2-CD technique to investigate the role evolutionary age and/or metallicity play 
in the development of the Be phenomenon is offered, and we provide evidence that other B-type objects
of very different nature, such as candidate Herbig Ae/Be stars may contaminate 
the claimed detections of ``Be stars'' via 2-CDs.  

\end{abstract}


\section{Introduction}

Classical Be stars are rapidly rotating, near main sequence B-type stars which 
show or have shown hydrogen Balmer emission \citep{jas81}.  It is widely accepted that
Be stars have gaseous, geometrically thin circumstellar disks \citep[for a review, see][]{por03}.   
As summarized in \citet{por03}, numerous mechanisms
have been proposed to explain how circumstellar disks may form around classical Be
stars, including the wind-compressed disk (WCD) model \citep{bjo93}, the magnetically torqued
wind-compressed disk model \citep{cas02}, and processes related to non-radial pulsations of the 
stellar photosphere \citep{riv01}, yet none have been completely successful in explaining the
observed phenomena.  Beginning with the work of \citet{str31}, it has often been speculated 
that the fundamental source of the 
Be phenomenon is tied to the rapid rotation of these stars: the canonical assertion is 
that classical Be stars rotate at v$_{eq}$/v$_{crit}$ $\sim$ 70-80\% of their critical 
velocity \citep{por96, por03}.  However,
recent theoretical work incorporating the effects of equatorial gravity 
darkening into studies of rotation rates \citep{tow04} suggests there is a degeneracy in the
measurement of rotation rates such that increasing v$_{eq}$ from 80\% to 100\% of the 
critical velocity has no effect on observed line widths.  Thus classical Be stars 
may actually be rotating at or near their critical breakup velocity.  Alternatively, \citet{cra05} 
suggest that a subset of classical Be stars might be rotating at sub-critical rates as low as 
0.4-0.6 v$_{crit}$. 

Studying stars from the Bright Star Catalog \citep{hof82} and its supplement 
\citep{hof83}, \citet{zor97} found that the mean frequency of Be stars 
with respect to normal B stars was $\sim$17\% in the Galaxy.  \citet{zor97} 
also found that the peak frequency occurred for spectral type B1, where
34\% of B stars were Be stars, and noted no difference in frequencies
between giant, dwarf, and subgiant type B stars.  It should be noted however that the
reliability of Bright Star Catalog spectral types is often questionable.  Early studies also
concluded that Galactic Be stars were present from the zero-age-main-sequence
to the terminal-age main sequence \citep{mer82,sle85}.  Due in part
to the advent of CCDs, systematic studies of the Be populations of star clusters and 
associations have expanded past our Galaxy to include
the Magellanic Clouds \citep{fea72,gre92,gre97,
die98,kel99,gre00,kel00,ols01}.  These extragalactic survey programs have typically used
2-color diagram (2-CDs) photometric techniques to identify the frequency of B-type stars
exhibiting excess H$\alpha$ emission (``Be stars'') relative to normal B-type stars in
stellar associations and clusters.  General trends in the fractional Be content of a 
cluster as a function of its age and/or metallicity have been observed, leading to 
suggestions that secondary mechanisms, besides rapid rotation, might influence the
development of Be circumstellar disks. 

\citet{mer82,gre97,fab00} and \citet{kel04} have found that the frequency of the Be phenomenon seems to peak in
clusters with a main sequence turn-off of B1-B2, leading to the suggestion that the Be phenomenon is
related to a star's evolutionary age.   \citet{fab00} suggested  
the Be phenomenon will start to develop only in the second half of a B star's main sequence lifetime, 
owing to structural changes in the star.  Specifically, \citet{fab00} noted that Be star-disk systems
 should start to appear in clusters 10 Myr old, 
corresponding to the mid-point main sequence lifetime of B0 stars \citep{zor97}, and their
frequency should peak in clusters 13-25 Myr old, corresponding to the mid-point main sequence 
lifetime of B1-B2 stars \citep{zor97}.  Curiously, \citet{kel99} claimed to detect 
numerous Be stars in the Small Magellanic Cloud (SMC) cluster NGC 346, 
which has an age between 2.6 Myr \citep{kud89} and 5 Myr \citep{mas89}; this result seems to conflict with
Fabregat \& Torrejon's assertion.  \citet{hil93} also suggested that numerous classical Be stars might be
present in the young cluster NGC 6611; however, it is possible that these objects are pre-main-sequence 
Herbig Ae/Be stars \citep{fab00}.  Numerous mechanisms have been proposed to explain this apparent 
evolutionary effect, including structural changes within the central star \citep{fab00}, an evolutionary spin-up of
the central star \citep{mey00, kel04}, and spin-up due to mass-transfer in binary systems \citep{mc05b}.

Several photometric studies \citep{gre92,maz96,gre97,mae99,kel04} have suggested that
the Be phenomenon may be more prevalent in low metallicity environments, based on comparisons
of the apparent fractional Be populations of Galactic, Large Magellanic Cloud (LMC), and 
SMC clusters.  IUE, HST, and FUSE observations
 \citep{gar85,bia96,ful00} show that stars tend to have lower wind velocities in metal poor environments.  
The WCD model \citep{bjo93} suggests that reducing terminal wind velocities would increase the 
likelihood that a B star experiencing
mass-loss would be able to retain this matter in the form of a circumstellar disk, although \citet{owo96} 
claim non-radial line forces may inhibit such a scenario.  \citet{ma99a} suggests that the 
increase in the frequency of Be stars in low metallicity environments is
related to the presence of a greater number of rapidly rotating stars in these locations,  
owing to a reduced coupling of the magnetic field and the young stellar object (\citealt{mae99}; see also 
\citealt{pen04}).

The accuracy of these initial interpretations is hampered by numerous
factors.  \citet{fab00} note that the photometric surveys used by many to
identify Be stars 
only detect Be stars with high levels of emission, and likely miss weak emitters.  
Furthermore, since Be stars are known variables which can experience
active and quiescent stages \citep{tel00,bjo02}, it is likely that some
Be stars in a cluster might be in a quiescent stage and hence not identified
as Be stars by these techniques.  This possibility has been demonstrated 
by various
followup investigations \citep{hum99,kel99}, which do not detect 100\% of
previously identified Be stars in certain clusters; additionally, a 
significant number of
previously unidentified Be stars are uncovered in these studies.  \citet{kel00}
note some spurious detections in photometrically identified Be stars located
in crowded cluster cores.  
 Even some spectroscopically identified Be stars
 \citep{maz96} have been shown to be spurious detections caused by the 
presence of diffuse background H$\alpha$ emission \citep{kel98}.  
The number of clusters with accurately determined
Be frequencies is still limited, especially regarding SMC clusters,
 thus small number statistics become important
when attempting to draw broad conclusions from the literature data set.

In this paper, we provide a significant increase in the number of extragalactic 
clusters whose candidate Be populations have been identified photometrically.  In section 2, we 
outline our observations and data reduction techniques.  In section 3, we detail our
identification of candidate Be stars in 8 SMC, 5 LMC, and 3 Galactic clusters.  We 
offer a detailed discussion of how these results may shed insight into the role 
age and/or metallicity play in the development of the Be phenomenon in Section 4.

\section{Observations and Data Reduction}

Our photometry was obtained at the Cerro Tololo Inter-American Observatory 
(CTIO)\footnote{The Cerro Tololo Inter-American Observatory is operated by the 
Association of Universities for Research in Astronomy, under contract with the National 
Science Foundation.} 0.9 m telescope using the direct imaging CCD, a 2048 x 2046 
multi-amplifier CCD, operated in the quad amplifier mode.  With the f/13.5
 secondary, we recorded data over a field of view of 
13$^{'}$.5 x 13$^{'}$.5 with a pixel scale of 0$\farcs$396 
pixel$^{-1}$.  We used CTIO's standard $3^{''}$ x $3^{''}$ B (
$\lambda_{center}$ = 4201\AA, $\delta \lambda$ = 1050\AA), V (
$\lambda_{center}$ = 5475\AA, $\delta \lambda$ = 1000\AA), R (
$\lambda_{center}$ = 6425\AA, $\delta \lambda$ = 1500\AA), and H$\alpha$ (
$\lambda_{center}$ = 6563\AA, $\delta \lambda$ = 75\AA) filters.  A
summary of our observations along with relevant cluster information is
given in Table \ref{allsum}.  Overscan correction, image trimming, 
bias correction, and flat fielding of the
data were achieved using standard IRAF\footnote{IRAF
is distributed by the National Optical Astronomy Observatories, which are
operated by the Association of Universities for Research in Astronomy, Inc.,
under contract with the National Science Foundation.} techniques.

To calibrate to the standard B,V,R system, we obtained photometry 
of 3-4 standard
fields, identified by \citet{lan92}, per night.  The observations were not
made under photometric conditions.  Aperture
photometry was performed for all standard stars using an aperture radius of
12 pixels.  
Transformations were derived from least-squares
fits to the following equations
\begin{displaymath}
 m_{b} = B_{0} + (B_{1}*X_{b}) + (B_{2}*(B - V))
\end{displaymath}
\begin{displaymath}
m_{v} = V_{0} + (V_{1}*X_{v}) + (V_{2}*(B-V))
\end{displaymath}
\begin{displaymath}
m_{r} = R_{0} + (R_{1}*X_{r}) + (R_{2}*(V-R)), 
\end{displaymath}
 where X is the airmass and the lowercase letters indicate
instrumental magnitudes in each filter.  The transformation coefficients
for each night are given in Table \ref{trans}.  Given our limited number of observations
of standard fields, we found it necessary to hold the the coefficient for the
airmass term constant to achieve a reasonable transformation.  As we are
only interested in differential photometric results, this should not 
adversely affect the analysis of our data.  Similarly, we did not calibrate our H$\alpha$ 
photometry to an absolute scale or calibrate for extinction as we were only 
interested in differential photometric results.

Due to the crowded nature of most of our clusters, unless otherwise
noted, final cluster photometry was obtained using standard IRAF point-spread
function (PSF) fitting techniques.  The FWHM of typical point sources varied in a non-linear
manner across the chip, thus we used a second order PSF function.  
Background sky levels were determined using mode statistics of an annulus of inner 
radius $4* FWHM$ having a width of $3* FWHM$.  Since the PSF photometry routine used an
aperture size, given by the average FWHM of stars in an image, which was
much smaller than the 12 pixel radius used for the photometry of the standard
fields, an aperture correction was applied.  

Achieving excellent absolute photometric accuracy was not an inherent goal of this study; 
nonetheless, for each of our fields of view we compared our measured photometry of several 
bright sources to published values (cross-listed in SIMBAD or \citealt{mc05b}) to search for 
signs of egregious systematic errors.  Our NGC 2186 photometry was the only data which exhibited 
any evidence of systematic errors; these data are consistently $\sim$1 V-band magnitude fainter than the 
photometric values cross-listed in SIMBAD.  These data were taken through clouds, thus the cause 
of the observed systematic error is likely related to the incomplete modeling of these effects via 
our observations of standard star fields.  We do not believe that these errors in the absolute 
photometry of NGC 2186 significantly affected our interpretation of the cluster's 
differential photometric behavior.

After completing the photometry measurements for all of our data, we organized our
 observations of each cluster into groups of short, medium, and deep
B,V,R,H$\alpha$ exposures (see Table \ref{allsum}).  Note that multiple listings
of H$\alpha$ exposures in a single grouping indicate that two separate
H$\alpha$ images were obtained to allow for co-addition of results and
hence lower errors.  Repetitive listings of individual targets in the short, medium, and 
deep exposures were identified and the observation having the lowest B-band error 
was selected for use in this study.  Finally, the photometric data were transformed into
the standard system using the transformation coefficients listed in Table \ref{trans}.

\section{Results}

\begin{figure}
\figurenum{1}
\epsscale{0.8}
\plotone{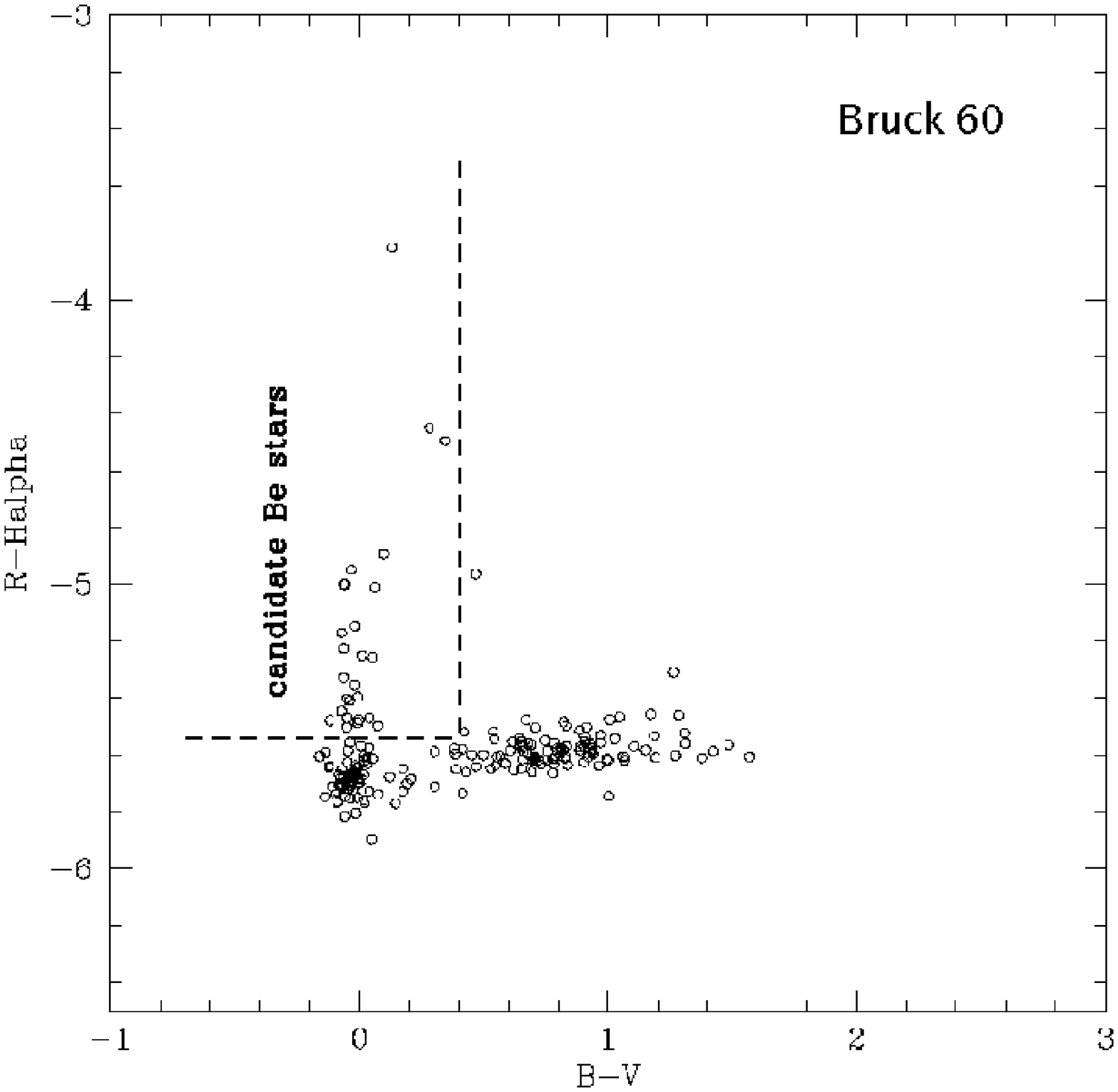}
\caption{Following the techniques established by e.g. Grebel and
Keller, we use 2 color diagrams to identify candidate Be stars in our
observed clusters.  Here we show an example 2-CD for the cluster Bruck 60.
The online version of this paper includes additional 2-CD
figures for the clusters: (F1.2) Bruck 107, (F1.3) Bruck 107 background population,
(F1.4) HW 43, (F1.5) NGC 371, (F1.6) NGC 456, (F1.7) NGC 458, (F1.8) NGC 460, (F1.9) NGC 465,
(F1.10) LH 72, (F1.11) NGC 1850, (F1.12) NGC 1858, (F1.13) NGC 1955, (F1.14) NGC 2027,
(F1.15) ELHC Field 2, (F1.16) ELHC Field 3, (F1.17) NGC 2186, (F1.18) NGC 2383, 
(F1.19) NGC2439. \label{2cd}}
\end{figure}

Following the techniques employed by Grebel, Keller, and others in the literature,
we used a simple 2 color diagram (2-CD) technique to identify the candidate
Be star populations in our clusters.  First, the B, V, R, and H$\alpha$ photometry of all 
stars in a cluster were plotted on a 2-CD, as illustrated in Figure \ref{2cd}.  
Normal blue main sequence
stars and most blue supergiants clumped in one section of this diagram, e.g. at
 $(B-V) = 0.0$ and $(R-H\alpha$) = -5.7 in Figure \ref{2cd}, 
while red main sequence and red supergiant stars associated in another region, e.g.  
at $0.3 < (B-V) < 1.5$ and $(R-H\alpha$) = -5.7 in Figure \ref{2cd}.  Blue stars 
which showed excess H$\alpha$ emission, e.g.
$(R-H\alpha) > -5.54$ in Figure \ref{2cd}, were designated  
\textit{candidate} Be stars.  Three criteria were used to delineate candidate Be stars from other 
astrophysical objects: (B-V) colors, (R-H$\alpha$) colors, and m$_{V}$ magnitudes. 

Our choice of maximum (B-V) colors delineating candidate Be 
stars from other astrophysical objects was based on the possible range of colors 
of classical Be stars in our clusters.  B-type stars of luminosity class III to V 
should exhibit unreddened (B-V) colors of -0.3 $<$ (B-V) $<$ $\sim$0 \citep{fit70,lan82}.  Classical Be disks 
can further redden the observed (B-V) colors by a few tenths of 
a magnitude \citep{sch78}; furthermore, 
interstellar reddening, as measured by the mean E(B-V) of our clusters (see Table \ref{allsum}), 
will also redden the aforementioned nominal color range.  However, this mean 
E(B-V) does not accurately characterize the reddening experienced 
by all cluster members.  For example, while \citet{ols01} cite a mean E(B-V) of 0.09 for LH 72, individual 
targets were observed to have E(B-V) values ranging from 0.04 to 0.22.  Thus some cluster objects may 
experience an additional reddening of about the same order as 
the quoted cluster mean E(B-V).  The 
(B-V) color cutoffs we used to identify candidate Be stars 
were (-0.3 + x + y + z) $<$ (B-V) $<$ 
($\sim$0 + x + y + z); where x represents the amount of 
reddening from Be disks ($\sim$0.2), y is 
the mean E(B-V) for each cluster, and z incorporates the possible affects of additional patchy interstellar 
reddening (1 times the quoted mean E(B-V)).  We stress that these cutoffs are only approximations 
meant to identify likely candidate Be stars; follow-up observations of these targets, especially the few which
lie on the borders of our chosen cutoffs, will be needed to confirm their status as bona-fide classical 
Be stars.  We will discuss individual (B-V) cutoffs for each of our clusters in the following subsections.

The choice of (R-H$\alpha$) colors delineating candidate Be stars from normal main sequence stars is 
somewhat arbitrary, and there is no systematic cutoff used throughout the myriad 
of literature studies which use 2-CDs to investigate classical Be populations.  \citet{mc05a} recently 
introduced an analysis technique based on synthetic photometry and empirical colors to alleviate such 
concerns, but this technique has yet to be widely adopted.  When assigning minimum 
(R-H$\alpha$) cutoffs to our clusters, we tried to ensure that our candidates showed noticeably more 
excess H$\alpha$ emission than likely main sequence stars which had slightly redder (B-V) colors than 
those expected for B-type stars.  Given the typical $(R-H\alpha)_{error}$ of less than 0.10  
and the sometimes considerable dispersion present in the blue and red
main sequence clumps, we suggest that our chosen $(R-H\alpha$) cutoffs dividing
blue main sequence stars from candidate Be stars represent a
conservative estimate of candidate Be star populations.  As expected, and as noted by 
other authors (e.g., \citealt{fab00}), such a
division will tend to exclude the detection of candidate Be stars which
exhibit low levels of Balmer emission.  

As a final constraint, we also ensured that the apparent magnitudes of candidate Be stars in each 
of our clusters coincided with those expected for B-type stars of luminosity class III-V.  Using the distances 
and reddening values for our clusters, we determined the upper (using B0III M$_{V}$ = -5.1, 
\citealt{lan82}) and lower (using A0V M$_{V}$ = 0.65, \citealt{lan82}) limits of expected apparent
magnitudes for candidate Be stars in our clusters.

Results for individual clusters are discussed below and summarized in Table \ref{photsum}.
Photometric data for each candidate Be star identified in this study, as well as astrometric 
coordinates accurate to $<$ 0.3 arc-seconds, are compiled in Table \ref{photdetails}.  
Note that the reliability of each candidate Be star detection in each 
cluster was
double-checked via inspection of each candidate's contour plot and
radial profile in multiple filters using IRAF.  A number of initial detections 
located within dense pockets of H II nebular emission were disregarded as we
believed their detection was the result of spurious background noise.
We also identified numerous objects which we classified as ``possible 
detections'' (see column 10 in Table \ref{photdetails}).  Due 
to either their location in dense H II regions or
possible nearby stellar contamination, we suggest these objects be
observed with secondary 
techniques, such as polarimetry or spectroscopy, to confirm or discount their
possible status as classical Be stars.  Note that data tabulated within parentheses in 
Table \ref{photsum} include both ``firm'' detections and ``possible'' detections of candidate
Be stars.

\subsection{Candidate Be Stars in SMC Clusters}
\subsubsection{Bruck 60}

\begin{figure}
\figurenum{2}
\epsscale{0.8}
\plotone{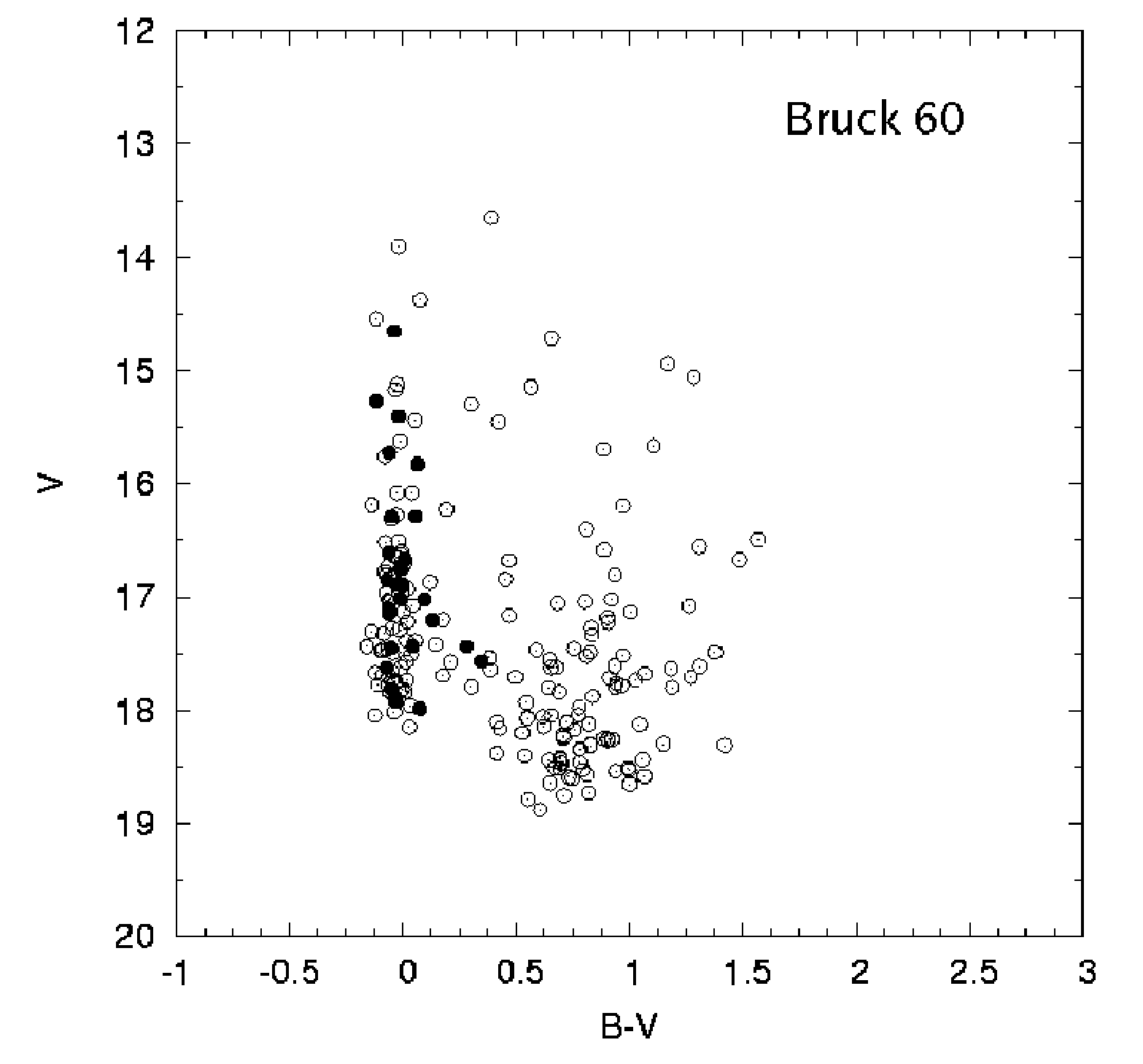}
\caption{The CMD of the SMC cluster Bruck 60 is presented here.  Filled circles
represent the location of candidate Be stars identified in this study.
Note that, as expected for classical Be stars, many candidates lie
slightly to the right of the main-sequence.  The online version of this
paper includes CMDs for the clusters:
(F2.2) Bruck 107, (F2.3) Bruck 107 background population,
(F2.4) HW 43, (F2.5) NGC 371, (F2.6) NGC 456, (F2.7) NGC 458, (F2.8) NGC 460, (F2.9) NGC 465,
(F2.10) LH 72, (F2.11) NGC 1850, (F2.12) NGC 1858, (F2.13) NGC 1955, (F2.14) NGC 2027,
(F2.15) ELHC Field 2, (F2.16) ELHC Field 3, (F2.17) NGC 2186, (F2.18) NGC 2383, 
(F2.19) NGC 2439. \label{cmd}}
\end{figure}

Based on the finder chart in \citet{kon80}, we defined the cluster size of Bruck 60 
to be a circle of diameter $3^{'}.0$.  Candidate Be stars were identified as
stars on the 2-CD (see Figure \ref{2cd}) with $ (B-V) < 0.35$, $(R-H\alpha) > -5.54$, 
and m$_{V}$ $>$ 13.9.  We identified 26 candidate Be stars based upon
this selection criteria, with 5 of these (Bruck 60:WBBe 7, 21, 23, 25, and 26) 
being classified as ``possible detections''.  The location of these 
candidate Be stars on the 
cluster color magnitude diagram (CMD) is given in Figure \ref{cmd}.  Note that the CMD and 2-CD for
Bruck 60 are presented here for illustration.  For subsequent clusters, the detailed CMDs and 2-CDs 
will be available in the online version of the Journal.  From an inspection of Bruck 60's CMD, as 
well as those for
our other clusters, one can see that our candidate Be stars sometimes lie redward of
the main sequence by (B-V) $\sim$0.1.  This likely indicates the presence of a small
amount of intrinsic reddening from these objects' circumstellar environments, as expected.

\subsubsection{Bruck 107}

\citet{kon80} defined membership in this cluster by a circle of
diameter $3^{'}.5$.  We identified 12 candidate Be stars from this cluster's 2-CD, using 
the selection criteria of $(B-V) < 0.20$, $(R-H\alpha) > -5.43$, and m$_{V}$ $>$ 13.9.  
\citet{kon80} also defined a ring with an inner
radius of $1^{'}.75$ and outer radius of $2^{'}.88$ as representative of
the background field star population.  Applying the same 2-CD selection criteria
as for the cluster population, we found 7 candidate Be stars in this suggested background
field population, with 1 of these detections (Bruck 107:WBBe 16) classified as a ``possible
detection''.  Note that in Table \ref{photdetails}, candidate Be 
stars labeled Bruck 107:WBBe 13-19 correspond to these field stars.

\subsubsection{HW 43}

\citet{kon80} defined cluster membership in HW 43 by a circle of diameter
$3^{'}.0$.  Seven stars with $(B-V) < 0.20$, $(R-H\alpha) > -5.50$, and m$_{V}$ $>$ 13.9 in this
cluster's 2-CD were designated as candidate Be stars. 

\subsubsection{NGC 371}

The cluster size for NGC 371 was taken from the work of 
\citet{hod85} and \citet{mas00}, 7$^{'}$.0 x 9$^{'}$.0.  The original cluster definition
in \citet{hod85} was a pseudo-potato shape whose major-minor axis
alignment was almost N-E.  We have used a 7$^{'}$.0 x 9$^{'}$.0 rectangle
oriented N-E to approximate this definition.  Because of completeness issues, we excluded all stars fainter
than V = 18.50 from our analysis.  129 candidate Be stars were
identified using the criteria $-0.40 < (B-V) < 0.50$ and $(R-H\alpha) > -5.45$.  Note that NGC 
371:WBBe 2 is 0.16 magnitudes too bright to be a B0III-B0V type star, given the distance modulus, 
R$_{V}$, and E(B-V) of the cluster.  However, follow-up polarimetric observations of this 
target \citep{wis05,wis06} clearly demonstrate that it is a classical Be star.   We therefore included it 
in the present study and adopted a m$_{V}$ magnitude cutoff of 13.8 for the cluster.  11 of these 
objects were designated as ``possible detections'': NGC 371:WBBe 19,
54, 60, 64, 71, 87, 90, 96, 122, 127, 129.  

\subsubsection{NGC 456}

The cluster size of 5$^{'}$.0 x 3$^{'}$.2 was based upon that used by
\citet{hi94a}.  23 candidate Be stars were identified using the criteria
$(B-V) < 0.5$, $(R-H\alpha) > -5.30$, and m$_{V}$ $<$ 14.5 and we consider 
1 of these candidates as a ``possible detection'', NGC 456:WBBe 21.

\subsubsection{NGC 458}

A circle of diameter 2$^{'}$.17 was used to define cluster membership 
\citep{mat02} in NGC 458.  We further excluded a small number of detections with
$(B-V)_{error} > 0.175$.  We assigned candidate Be star status to
all objects with $-0.3 < (B-V) < 0.2$, $(R-H\alpha) > -5.42$, and m$_{V}$ $<$ 13.9 resulting
in the identification of 30 objects. 
 2 ``possible detections'' were included in this total, NGC 458:WBBe 5 and 19.

\subsubsection{NGC 460}

The dimensions of this cluster, $4^{'}.6$ x 6$^{'}.0$, were adopted from \citet{hi94a}.
 21 candidate Be stars with $(B-V) < 0.2$, $(R-H\alpha) > -5.40$, and m$_{V}$ $>$ 14.1 were
identified from this cluster's 2-CD, with 7 identified 
as ``possible detections'', NGC 460:WBBe 1, 9, 13, 14, 19, 20, and 21.

\subsubsection{NGC 465}

A cluster size of $5^{'}.3$ x $5^{'}.3$ was assumed based upon the definition
of \citet{hi94a}.  11 candidate Be stars with $(B-V) < 0.25$, $(R-H\alpha) 
> -5.40$, and m$_{V}$ $>$ 14.0 were identified in this cluster's 2-CD.

\subsection{Candidate Be Stars in LMC Clusters}

\subsubsection{LH 72}

The unique cluster shape designated by \citet{luc72}, kindly supplied to us by
K. Olsen (2003, private communication), was used to define LH 72 cluster
membership.  We eliminated all stars with $V > 18.7$ from our analysis owing to completeness issues.
Candidate Be stars were defined as objects with $(B-V) < 0.4$, $(R-H\alpha) 
> -5.60$, and m$_{V}$ $>$ 13.7 in the cluster's 2-CD, resulting in 50 detections.  
Note that 11 of the 50 candidate Be
stars were designated ``possible detections'': LH 72:WBBe 2, 15, 18, 20, 23, 
37, 38, 41, 43, 46, and 48.  

The Be population of this cluster
and its nearby vicinity was previously investigated \citep{ols01}, with
a fractional Be population of ``at least 10\%'' reported by these authors.  We have attempted 
to correlate these previously suggested Be
stars, using the x and y pixel coordinates kindly supplied to us by
K. Olsen (2003, private communication), with the candidate Be stars identified
in this study.  With the exception of the two Be stars for which \citet{ols01} 
listed RA and Dec coordinates, we were unable to make firm correlations
between the data sets, and we suggest two explanations.  As discussed in the
introduction, Be stars are variable stars known to periodically enter
quiescent phases in which they lose most or all of their circumstellar disks.
Thus it is not unexpected that we might identify a different set of candidate Be
stars than those identified from a previous epoch.  LH 72 also resides in a region of significant
H$\alpha$ nebulosity.  It is likely that both \citet{ols01} and the present
study mis-identified a small number of candidate Be stars due to
 spurious detections induced by
this diffuse, heterogeneous background.  Each study also used
a different method of accounting for this diffuse background emission; hence, one might 
expect that different types of mis-detections might be present in our study as 
compared to those present in \citet{ols01}.

\subsubsection{NGC 1850}

We defined the cluster dimensions of NGC 1850 to be $3^{'}.8$ x $3^{'}.8$.  \citet{val94} 
suggested that this region actually contains 
three distinct groupings of stars, NGC1850, NGC1850A, and H88-159, which
are in slightly different evolutionary stages.  Given the extremely crowded
nature of the center of the field and the quality of our images, we 
were unable to accurately separate these cluster components and thus have considered
the entire region as a single unit.  Additionally, we have excluded all stars
from our analysis which have $m_{V} > 18.50$, m$_{V}$ $<$ 14.0, 
$(R-H\alpha)_{error} > 0.10$, and $(B-V)_{error} > 0.18$.

Inspection of the cluster's 2-CD reveals considerable noise, 
which we
attribute to crowding effects.  We therefore imposed very conservative
candidate Be criteria of $-0.35 < (B-V) < 0.40$ and $(R-H\alpha) > -5.30$.
  Using
these guidelines we identified 92 candidate Be stars, of which we classified
6 as ``possible detections'': NGC 1850:WBBe 4, 29, 52, 57, 58, and 84.

\subsubsection{NGC 1858}

The parallelogram cluster shape used by \citet{val94} was used to define
cluster membership in our observation of NGC 1858.  We excluded all objects $V > 19.0$
from our analysis.  39 candidate Be stars were identified using the criteria 
that $-0.3 < (B-V) < 0.4$, $(R-H\alpha) > -0.30$, and m$_{V}$ $>$ 13.9 in this cluster's
2-CD, with 4 candidates 
classified as ``possible detections'': NGC 1858:WBBe 9, 13, 24, and 35.

\subsubsection{NGC 1955}

The cluster size of $4^{'}.2$ x $3^{'}.5$ was based upon that used by
\citet{hi94a}.  Candidate Be stars were identified using the criteria
$(B-V) < 0.25$, $(R-H\alpha) > -5.50$, and m$_{V}$ $>$ 13.7 leading to the detection of 24
targets.  3 of these targets were classified as ``possible detections'': 
NGC 1955:WBBe 9, 14, and 24.  

\subsubsection{NGC 2027}

The cluster dimensions of $9^{'}.0$ x $5^{'}.0$ were adopted from 
\citet{luc70}.  We excluded all stars with $V > 19.0$ from our analysis.
Based on the selection criteria of $(B-V) < 0.3$, $(R-H\alpha) > -5.40$, and m$_{V}$ $>$ 13.6 
we identified 46 candidate Be stars, of which 3 were classified as 
``possible detections'': NGC 2027:WBBe 44, 47, and 48.  

\subsubsection{ELHC Fields 2,3}

\citet{lam99} and \citet{dew02} identified 21 EROS LMC Herbig Ae/Be (HAeBe)
 Candidates
(ELHCs) by searching the EROS2 mircolensing photometry database for irregular
variables which exhibited similar properties to Galactic Herbig Ae/Be stars.  We observed two LMC fields
which contained many of these ELHCs to determine
if they exhibited excess H$\alpha$ emission, as would be expected if 
they were truly intermediate-mass pre-main-sequence objects. 

Since the ELHCs were dispersed across a significant portion of the LMC, we
analyzed the stellar content of the full $13^{'}.5$ x $13^{'}.5$ field of view
of both of the LMC fields we observed.  We limited our analysis to stars
with m$_{V}$ $>$ 13.4, $m_{V} < 19.0$, $(B-V)_{error} < 0.15$ and $(R-H\alpha)_{error} < 0.15$.
For the ELHC field 2 image, we assigned the designation ``candidate
emission-line stars'' to 183 stars having $(B-V) < 0.50$ and
$(R-H\alpha) > -5.35$.  We detected 4 of the 5 previously designated
 ELHCs in our field of view, ELHC 3, ELHC 4, ELHC 7, and ELHC 19;
however, we note that 2 of these detections, ELHC 3 and ELHC 4, were very close
to our minimum detection criteria.  The star ELHC 20 \citep{dew02}, which we
observed to have (R-H$\alpha) = -5.61$, was not detected as a candidate 
emission-line star on our 2-CD.  Figure 2.15, available in the online edition of this Journal, 
presents the CMD of this field of view; the four detected ELHC stars are plotted as 
large circles, while the remaining 179 ``candidate emission-line stars'' are not identified by 
any special type of symbol.

In our ELHC field 3 image, 153 ``candidate emission-line stars'' were
identified using the criteria $(B-V) < 0.50$ and $(R-H\alpha) > -5.20$.
We detected 5 of the 7 previously designated ELHCs in this field of view,
ELHC 1, ELHC 6, ELHC 8, ELHC 12, and ELHC 13.  ELHC 5, with $(R-H\alpha) = -5.39$,
was not detected as a candidate on our 2-CD.  We found ELHC 11 to be composed of two stellar
components separated by $\sim$2$^{''}.4$.  The $(R-H\alpha$) magnitude of each
of these components, -5.55 and -5.77, lay outside of our selection criteria
for candidate emission-line objects.  Unfortunately we could not comment on
the status of ELHC 9, which was also in this field of view, as it was centered
on a significant cluster of bad columns on the CCD.  Figure 2.16, available in the online edition 
of this Journal, presents the CMD of this field of view.  The five detected ELHC stars are 
plotted as large circles in this figure, while the remaining 148 ``candidate emission-line stars'' 
are not identified by any special type of symbol.

\subsection{Candidate Be Stars in Galactic Clusters}

We also examined the fractional Be content of three Galactic clusters.  Given the
diffuse nature of these open clusters, we used aperture photometry to
analyze these data.

\subsubsection{NGC 2186}

We observed this cluster, as in spite of its relatively young age, the 
WEBDA database \citep{mer03} indicated that no one had identified its
population of classical Be stars.  We defined
the cluster size as a $5^{'}.8$ x $5^{'}.2$ box which corresponds
to the field considered by \citet{mof75}.  Recall from Section 2 that our NGC 2186 data seem 
to suffer from systematic photometric errors which make all objects appear $\sim$1 magnitude 
fainter than published photometry of the cluster.  We have not corrected our 
data for these systematic effects; however, we do not believe these errors significantly affect 
our differential photometric results.  Candidate Be star status was 
assigned to 5 stars with $(B-V) < 0.40$, $(R-H\alpha) > -4.80$, and m$_{V}$ $<$ 14.0.   
Note that in the absence of our systematic photometry errors, we would have applied a 
m$_{V}$ cutoff of 13.0 for this cluster.

\subsubsection{NGC 2383}

At the time of our observations, no one had identified the Be population of NGC 2383
\citep{mer03}; subsequent to this time \citet{mc05b} have published a photometric study of the 
cluster in which they identified two Be stars, labeled as stars \# 11 and \# 341 in their survey.  There
is considerable disagreement in the literature concerning the age of 
NGC 2383 and the spectral classification of some of its members.
\citet{lyn95} claim the cluster's age is log (t) = 7.4, while \citet{sub99}
claim a much older age of log (t) = 8.6.  The spectral classifications
given in \citet{sub99} also differ significantly from previous classification
attempts, e.g. they classify their star S1 as an A3 I object versus a previous
classification of B0 III.  

\citet{sub99} defined cluster membership to extend to a diameter of $5^{'}.0$,
which we also adopted.  We identified 3 candidate Be stars on the cluster's 2-CD
using the criteria $(B-V) < 0.40$ and $(R-H\alpha) > -5.78$.  As there is a large uncertainty 
in the distance to this cluster, we did not define an upper m$_{V}$ cutoff when determining this  
cluster's candidate Be star population.  For similar reasons, we have not tabulated 
the number of B-type main sequence stars for this cluster in Table \ref{photsum}.
Comparing our dataset to that of \citet{mc05b}, one of their Be star detections (star \# 341) was 
detected in the present study as NGC 2383:WBBe 1 while their other Be star detection (star \# 11) 
was not detected by us.  Our other two candidate Be star detections, NGC 
2383:WBBe 2 and 3, were not flagged as Be stars by \citet{mc05b}.

\subsubsection{NGC 2439}

We adopted a cluster diameter of 10$^{'}.0$ based upon the cluster map of
\citet{whi75}.  We further restricted our analysis to stars with 
$(R-H\alpha)_{error} 
< 0.1$.  Note that we strongly suspect that our group of
shortest exposures of this cluster have uncertain exposure
times due to the limited fastest shutter speed of the CTIO 0.9m.  As a result, the
few cluster stars we report from this set of exposures appear to be bluer
by $\sim$0.5 than the other exposure sets.   

\citet{sle85} and references therein note the presence of
5 Be stars in the cluster, labeled as stars 6, 69, 75, 81, and 303.  We identified 4 of these 5 
previously known Be stars, numbers 6, 69, 81, and 303, based on a 2-CD criteria of
$(B-V) < 0.5$, (R-H$\alpha$) $>$ -5.80, m$_{V}$ $>$ 9.3, and m$_{V}$ $<$ 15.0.  Star number 41
\citep{whi75} also appeared as a candidate Be star in our data, although it
was not previously detected as such.  In contrast, the previously identified
Be star number 75 in the study by \citet{sle85} was not detected as such
by our data set.  More recently, \citet{mc05b} used 2-CDs to identify 6 Be and 7 
candidate Be stars associated with NGC 2439.  The present study has confirmed 
4 of 6 of the Be detections reported by \citet{mc05b} (WBBe 1 = their no. 58, 
WBBe 2 = their no. 24, WBBe 3 = their no. 101, WBBe 4 = their no. 22) and 1 of the 
7 candidate Be detections reported by \citet{mc05b} (WBBe 5 = their no. 41). 
The variability in the number of detections and non-detections amongst these surveys is 
likely due in part to the variable nature of classical Be stars, and these comparisons provide 
support for the idea that the identification of candidate Be stars may be subject to such 
variability issues.  Additional factors such as 
differences in the defined size of the cluster (i.e. \citealt{mc05b} versus the present study) 
also influence the reported fractional content of Be stars.

\section{Discussion}


In order to better probe the role of evolutionary age and/or metallicity may play in the development
of the Be phenomenon, we approximated rough spectral types for our candidate Be stars on the 
basis of rough magnitude bins associated with spectral ranges.  We first transformed
the apparent magnitudes of our candidates to the absolute scale, using
the standard equation m$_{V}$ - M$_{V}$ = 5 log d - 5 + A$_{V}$, where 
A$_{V}$ = R$_{V}$ * E$_{(B - V)}$.  For LMC clusters we used a distance 
modulus of 18.5 \citep{wes90} and 
R$_{V}$ = 3.41 \citep{gor03}, while for SMC clusters we adopted a distance
modulus of 18.9 \citep{wes90} and R$_{V}$ = 2.74 \citep{gor03}.  For our
Galactic clusters we assumed R$_{V}$ = 3.1 and used a distance modulus of 
11.31 for NGC 2186 \citep{mof75} and
13.24 for NGC 2439 \citep{whi75}.  E$_{(B - V)}$ values for most clusters are listed
in Table \ref{allsum}.  Due to the lack of available published reddening values for
Bruck 60, Bruck 107, and HW 43, we assigned these clusters the mean reddening
value for the SMC, 0.037, as determined by \citet{sch98}.  Following the work of 
\citet{gre97}, we used the calibration for main sequence stars given in 
 Table 5 of \citet{zor97} to transform the absolute magnitude of each of our 
candidate Be stars into a spectral type.  As discussed
by \citet{gre97}, a large uncertainty is associated with such a transformation
technique; hence, our estimated spectroscopic classifications should be considered only crude
approximations.  

We summarize these rough spectroscopic classifications for our LMC and SMC clusters
in Table \ref{photspectypesum}.  
In column 2 of Table \ref{photspectypesum}, we assigned an age label to each cluster given by: 
5 Myr $<$ Age $<$ 8 Myr = very young (vy), 10 Myr $<$ Age $<$ 25 Myr = young (y),
and 32 Myr $<$ Age $<$ 158 Myr = old (o).    Note that we classified LH 72 as ``very young'' (vy) 
although a range of ages have been suggested for this cluster 
(see Table \ref{allsum}).
Given the suggested multiple epochs of star formation in NGC 1850 (see 
Section 3.3.2), we designated this cluster as ``old?'' (o?).  We have included all photometrically 
identified candidate Be stars in Table \ref{photspectypesum}, including those classified as
``possible detections'' in Table \ref{photsum}.

We have combined the results of Table \ref{photspectypesum} into two bins: 
``early-type'' candidate Be stars, corresponding to rough spectral types B0-B3, and ``later-type'' 
candidate Be stars, corresponding to rough spectral types B4-B5.  The results of this binning are given
in Table \ref{earlylate}, and we suggest that such averages should mask 
the general level of uncertainty associated with 
the spectral types we have assigned.  Clusters with fewer 
than 20 B-type main sequence objects in either of these spectral bins were excluded from
consideration to lessen the effects of small number statistics.  The standard deviation 
errors quoted in Table \ref{earlylate} were calculated assuming the data followed a
binomial distribution.  We have also averaged the fractional Be content, binned to early- and later-type stars, 
of clusters having like age and metallicity properties, as summarized in 
Table \ref{agezstats}.  The uncertainties quoted in this table merely represent the
propagation of the previously calculated standard deviations, i.e. 
$\sigma$ = N$^{-0.5}$ ($\sigma_{i}^{2}$ + $\sigma_{i+1}^{2}$ + ...)$^{0.5}$.

\subsection{Metallicity Effects}

\begin{figure}
\figurenum{3}
\epsscale{0.8}
\plotone{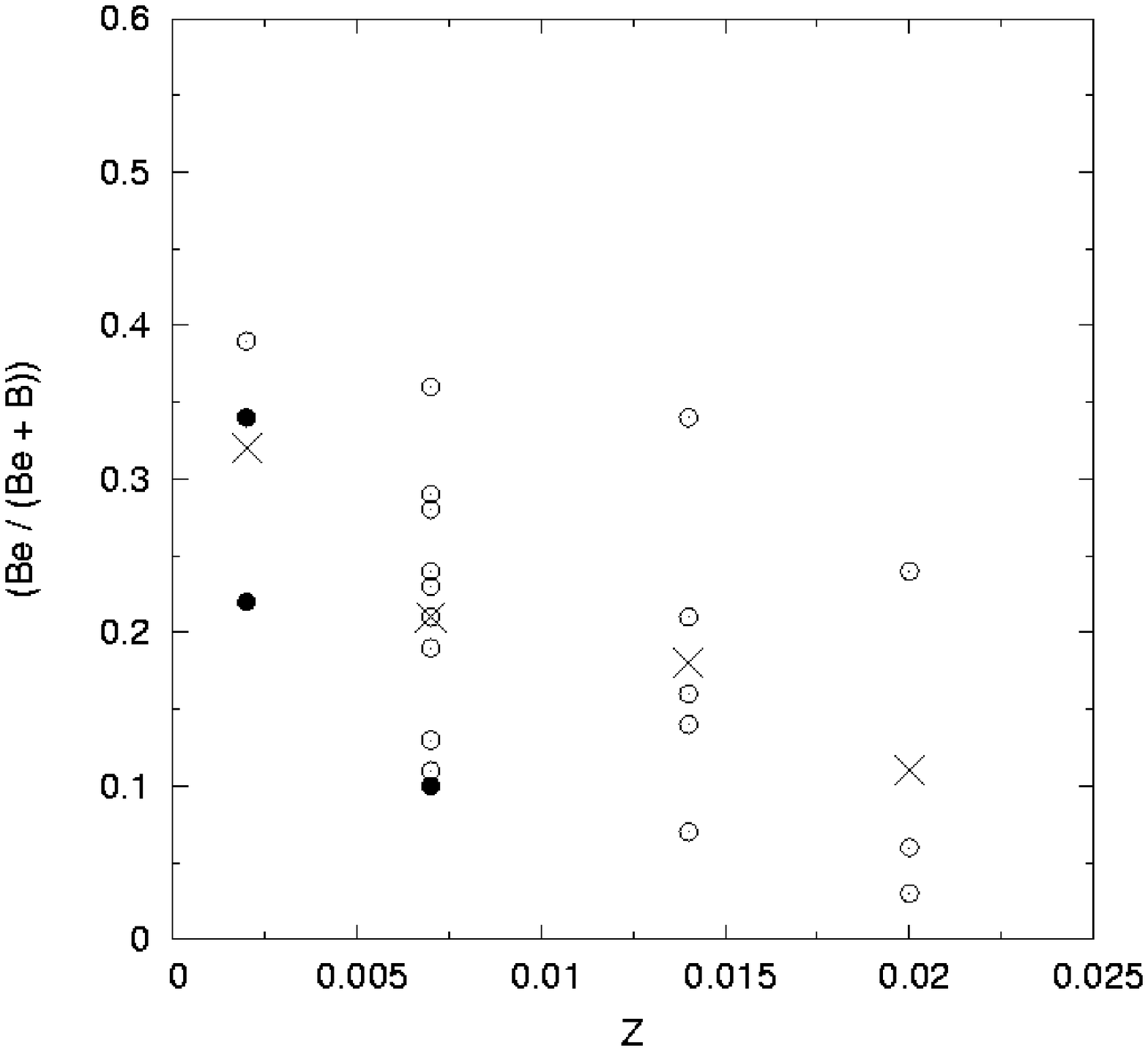}
\caption[f3.eps]{The fractional candidate Be star 
content of B0 - B3 type objects in ``young'' clusters with ages ranging from
7.0 $<$ log(t) $<$ 7.4 is shown as a function of cluster metallicity.  Data from 
the present study, \citet{kel99}, and \citet{mae99} are plotted, excluding clusters which had fewer than 
20 early-type (B + Be) stars.  As discussed in Section 4.1, we have followed the practice of 
\citet{mae99} and assigned average metallicity values for our SMC clusters (z = 0.002), 
LMC clusters (z = 0.007), 
Galactic clusters located exterior to the Solar location (z = 0.014), and Galactic clusters located interior to the 
Solar location (z = 0.020).  The 
filled circles represent data presented in this study, the open circles represent literature data, and the 
large crosses represent the average of all cluster data within a metallicity bin. \label{midagez}}
\end{figure}

We first examined our dataset to identify any trends in the fractional candidate Be content
as a function of metallicity.  Following the work of \citet{mae99}, we first considered 
the fractional early-type (B0-B3) Be content of ``young'' clusters having 
ages of 7.0 $<$ log (t) $<$ 7.4.  Data matching these criteria in Table \ref{earlylate} 
are plotted in Figure \ref{midagez}, where open circles correspond to literature data, 
filled circles correspond to data presented in this study, and crosses represent the
average of all clusters' fractional Be content within a metallicity bin.  It is possible to find a wide 
range of reported metallicities for many of the individual clusters presented in this survey, as noted by 
\citet{mae99}; hence, following their example, we have chosen to represent all SMC clusters with one 
average metallicity and all LMC clusters with another average metallicity value.  We also follow the 
practice of \citet{mae99} of assigning single, average metallicity values to both Galactic clusters located 
exterior and interior to the Solar location (see Figure \ref{midagez}).  We recognize that this is an 
oversimplification, but given the large scatter in the current available values for individual cluster
metallicities, it seems the best approach at present.

\begin{figure}
\figurenum{4}
\epsscale{0.8}
\plotone{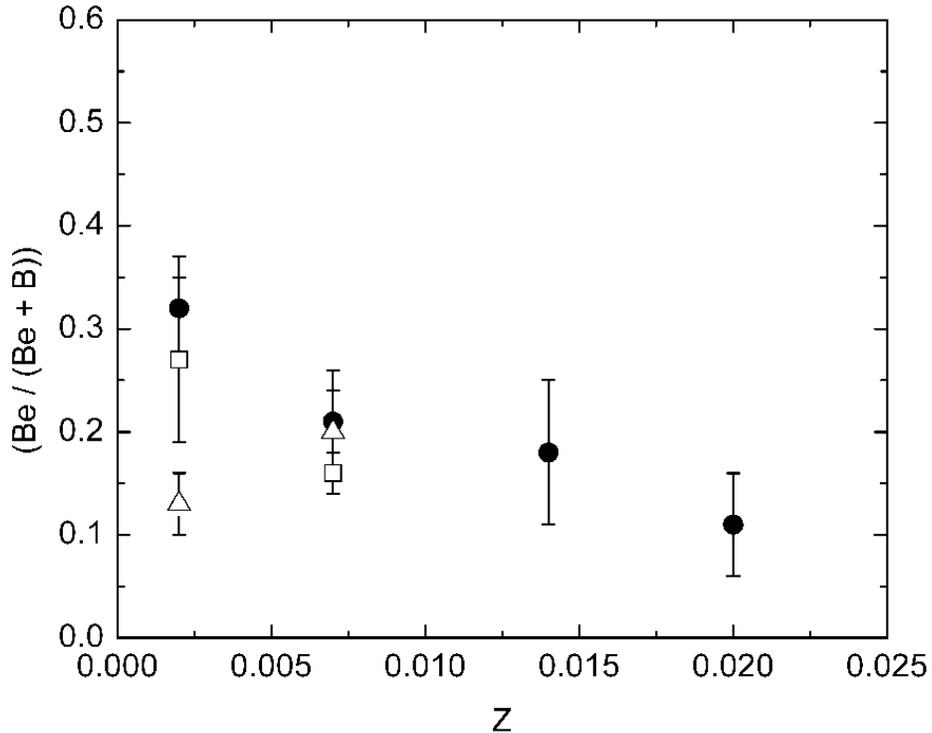} 
\caption[f4.eps]{The fractional candidate 
Be content of clusters of various age denominations were averaged to examine
systematic trends with metallicity.  Open triangles correspond to ``very young'' clusters, closed circles correspond to ``young'' clusters, and open squares correspond to ``old'' clusters.  As discussed 
in Section 4.1, we have followed the practice of \citet{mae99} and assigned average metallicity 
values for our SMC clusters (z = 0.002), LMC clusters (z = 0.007), Galactic clusters exterior 
to the Solar location (z = 0.014), and Galactic clusters interior to the Solar location (z = 0.020). 
\label{averagedz}}
\end{figure}

From Figure \ref{midagez}, it is clear that there exists a wide range of Be / (B + Be) ratios 
within each metallicity bin.  The systematic errors inherent in the use of the 2-CD technique 
undoubtedly contribute to this scatter; differences in the environmental properties of specific individual 
clusters also might influence the observed scatter of fractional Be content.  However, it is clear 
that a trend in the fractional Be content 
with metallicity exists.  This trend is better seen in Figure \ref{averagedz}, where the filled circles 
represent the average fractional Be content of ``young'' clusters.  The average fractional Be content of 
SMC metallicity clusters is $>$ 2$\sigma$ higher than the average Galactic (interior or exterior) value, 
as documented in Table \ref{agezstats}.

While Figure \ref{averagedz} shows the same general trend depicted by \citet{mae99}, we believe
that the improved statistics provided by the present study significantly strengthens the claim of 
a trend with metallicity.  The trend claimed by \citet{mae99} was based upon the observation of
one cluster in the SMC metallicity bin of z=0.002.  The present study's use of 4 SMC 
clusters reduces the average fractional Be content within this metallicity bin from 39\% \citep{mae99} 
to 32\%.  We must note that our results are not inconsistent with the general conclusion reached by 
\citet{mc05b}, who found no evidence of a metallicity trend in their study of Galactic clusters.  As seen 
in Figure \ref{averagedz}, we also observe little convincing evidence of a trend over the smaller metallicity 
range probed by Galactic clusters; however, evidence of a metallicity trend develops when one 
extends surveys to include broader regions of metallicity space, such as the LMC and SMC.

If one assumes that the fractional Be content of clusters depends on metallicity, then such 
a trend also should be apparent when examining clusters in age
denominations other than ``young'' clusters.  We have plotted the average fractional 
early-type Be content of ``very young'' 
clusters, depicted as open triangles, and ``old'' clusters, depicted as open squares, in 
Figure \ref{averagedz}.  Although the quality of available data for these age groups is
severely limited, and no data exist for Galactic clusters, we suggest that the ``very young'' and ``old'' cluster
data plotted in Figure \ref{averagedz} are not inconsistent with the Be phenomenon being more prevalent
in low metallicity environments.  We do recognize, however, that there is still some controversy 
over the metallicity effect.  For example, the results of \citet{mar06} find a Be fraction in the LMC 
cluster NGC 2004 which is about the same as the standard Galactic value of 17\%.  
To resolve these lingering uncertainties, we strongly suggest
that additional observations of clusters in these extreme age ranges must be made, spanning a wide range
of metallicities, in order to provide the statistics necessary to more definitively identify any trend present.

\subsection{Evolutionary Age Effects}

Inspection of Tables \ref{photspectypesum}, \ref{earlylate}, 
and \ref{agezstats} clearly reveal that we have found a substantial number of candidate Be stars 
in clusters ranging in age from 5 Myr to 8 Myr.  From Table \ref{photspectypesum}, it is clear that 
these young clusters not only contain B0 type candidate Be stars, but
they also have a significant number of later-type candidate Be objects.  If the Be phenomenon
develops in the second half of a B star's main sequence lifetime, 
we would not expect to find B0 type stars, let alone B4-B5 type stars, in these very young
clusters.  Furthermore, the average fractional candidate Be content in these very young clusters, detailed
in Table \ref{agezstats}, is similar to the nominal frequency of the Be phenomenon in the Galaxy of 
17\% \citep{zor97}.  

Clearly our present understanding of how the Be phenomenon is related 
to a star's evolutionary age would be altered if we could establish that these very young \textit{candidate} 
Be stars were not remnant pre-main-sequence star-disk systems.  The similarities 
between the fractional candidate Be population found in these very young 
clusters with respect to the nominal frequency of the Be phenomenon observed in the Galaxy 
is interesting.  Classical Be stars of such a young age clearly would 
not have spent enough time on the main sequence to significantly spin-up via the mechanism proposed by 
\citet{mey00}, or via mass transfer in a binary system \citep{mc05b}.  Since it is widely accepted that the 
Be phenomenon is
inherently tied to rapid rotation, our data seem to suggest that 
stars emerging from their pre-main-sequence phase must possess a wide range of rotational 
velocities; namely, a significant number of objects must be rotating near their critical breakup
velocities at the zero-age-main-sequence (ZAMS).  We will explore the true nature of many of these 
extremely young candidate Be stars in a later paper.  
 
From Table \ref{agezstats}, it is clear that the fractional early-type Be content of ``young'' clusters 
is significantly higher than the nominal frequency of the Be phenomenon (17\%); furthermore, the 
fractional later-type Be content of these clusters is significantly lower than the early-type content.  
Examining the fractional early- and later-type 
Be content of ``old'' clusters in Tables \ref{earlylate} and \ref{agezstats}, we also find evidence that the
Be phenomenon in these clusters is more prevalent than the nominal value of 17\%.  It is unclear from our 
dataset whether there is any difference between the frequency of early-type and later-type Be stars in
these old clusters.  Our detection of early-type objects in ``old'' clusters is very curious, and we 
can not find a simple explanation in the literature which explains why such objects 
would be detected.  The slight enhancement 
of the fractional later-type Be content in our old clusters 
and the much larger enhancement of the fractional early-type Be content in our young clusters supports 
previous suggestions that the Be phenomenon is more prevalent in the later stages of a B star's main 
sequence lifetime \citep{mer82,gre97,fab00,kel04,mc05b}.  Recall however, that the results 
from our very young clusters indicate that such an 
evolutionary effect \textit{may not} be the sole mechanism responsible for the development of the
Be phenomenon.

We now consider the mechanisms which might be responsible for the observed evolutionary effect 
in the Be phenomenon.  The detection of a sizable fraction of candidate Be stars in 
clusters younger than 10 Myr suggest that stars emerging onto the ZAMS
might display a wide range of rotational rates, including stars which are rotating near their critical breakup
 velocity, hence 
exhibiting the Be phenomenon.  The theoretical models of \citet{mey00} and \citet{mae01} illustrate 
that the ratio of angular velocity to critical angular velocity ($\Omega$ / $\Omega_{critical}$) 
steadily increases throughout the main sequence lifetime of early-type B stars, which 
these authors suggest might 
explain why the Be phenomenon is more prevalent in the later part of a B star's main sequence
lifetime.  While we agree that the general notion of an evolutionary spin-up provides a reasonable 
explanation for most of the observational properties documented here and in the literature, we 
also consider some of other implications of Meynet's and Maeder's models.

Table \ref{lifetimes} lists the main sequence lifetimes predicted by the models of \citet{mey00} and 
\citet{mae01} for rotating and non-rotating stars.  Interestingly, stars starting with an enhanced rotation rate 
on the ZAMS will experience a longer main sequence lifetime than stars of similar mass which start with a
small or null rotation rate.  Recall that the fundamental underlying mechanism believed to differentiate 
classical Be stars from 
``normal'' B-type objects is rapid rotation.  We speculate that the different main 
sequence lifetimes experienced by rapidly rotating versus slowly rotating B-type stars might influence 
the observed enhancement of the Be phenomenon in the later stages of a star's main sequence lifetime.
Consider a SMC cluster of age 25 Myr, which Table \ref{lifetimes} suggests will only have B3 or later-type 
``normal B stars'' remaining on the main sequence, assuming they started on the ZAMS as slow rotators.  In addition 
to B3 or later-type rapidly rotating classical Be stars, 
B2 type classical Be stars will also still reside on the main sequence when the cluster is 25 Myr old.  Hence, 
the fractional Be content derived for this cluster via the 2-color diagram technique will 
not be comparing Be to normal B stars of a similar range of spectral types.  The additional presence of B2 type
Be stars in this cluster might inflate the observed fractional Be content.

Since clusters younger than 10 Myr old should still have their entire B0-B9 sequence present on the
main sequence, regardless of whether one considers rapidly or non-rapidly rotating stars, we expect no
enhancement of the Be phenomenon due to this effect.  Conversely, we would expect that the expansion of
main sequence lifetimes due to rotation will greatly affect clusters in the 16-26 Myr age range.  In 
these clusters, ``normal'' B1-B2 stars 
which were slowly rotating when they reached the ZAMS would have evolved off of the main sequence.  
B1-B2 type classical Be stars, which were rotating much more rapidly when they reached the ZAMS, 
will still be observable on the main sequence, hence B1-B2 and B3-later type Be stars will be detected
via 2-CDs.  Because the Be phenomenon is most prevalent amongst B1-B2 type stars, one might expect
a very large enhancement of the Be phenomenon to be detected in these clusters.  When one considers
the fractional Be content of older clusters, the Be phenomenon would likely still be enhanced by this effect, 
but to a lesser degree than clusters 16-26 Myr old, as the prevalence of the Be phenomenon is lower in 
later spectral types.

Our data are consistent with this speculative scenario.  No strong enhancement in the Be 
phenomenon is seen in our 
very young clusters, while moderate and strong enhancements in the Be phenomenon are seen in our old and
young clusters respectively.  From Table \ref{agezstats}, one might question why we observe early-type 
objects in our ``old'' clusters.  We would expect that most slowly rotating early-type stars should have
evolved off of the main sequence in these clusters.  A limited number of early-type classical Be stars 
might still reside on the main sequence, owing to the extended main sequence lifetime afforded to 
rapid rotators.  It is a little more difficult to explain the nature of early-type non-Be stars detected in these
clusters; however, it is possible that these objects are blue stragglers.  Note that the binary scenario for 
the formation of Be stars predicts that some should appear as blue stragglers \citep{pol91}; hence, if 
``normal'' B-type blue stragglers populate these clusters, some of the clusters' Be population may have 
their origin linked to binarity, not the suggested extension of main sequence lifetimes.  It is clear that 
additional observations are needed to investigate the nature of both 
the early-type candidate Be stars and normal B stars in such ``old'' clusters.

Differentiating between an enhancement in the Be phenomenon which is due to the spin-up of stars as 
they evolve along the main sequence versus an extension of the main sequence lifetime in rapidly
rotating stars will likely be a difficult task.  Of course, it is entirely possible that both mechanisms 
could play a role in producing the observed enhancements.  One could examine the statistical distribution of rotational 
velocities of early-type versus later-type B stars in clusters spanning a range of ages to look for 
evidence of a systematic spin-up of the overall distribution with age.  Conversely, if one 
could find numerous classical Be stars 
in a cluster having earlier spectral types than all of the normal main sequence stars present, this 
would provide evidence that the extended main sequence lifetimes of rapid rotators might contribute to
the observed enhancements of the Be phenomenon.  In practice, measuring the spectral types of 
classical Be stars to high accuracy is difficult, so this later type of observational undertaking would likely
be challenging.

\subsection{Galactic Clusters}

We also used the calibration of \citet{zor97} to assign crude spectral types
to the candidate Be stars we identified in our Galactic clusters (see Table 
\ref{galacticspectypes}).  Due to the large uncertainty
in the distance to NGC 2383, we have not 
attempted to convert the observed apparent magnitudes of this cluster's candidate Be
stars into spectral types.  Recall that our NGC 2186 photometry includes systematic errors which 
make our targets appear $\sim$1 magnitude too faint; to correct for these systematics, 
we have applied a 1 magnitude offset to our data 
to compute the spectral types listed in Table \ref{galacticspectypes}.  As discussed in 
Section 3.3.3, we detected 4 of 5 previously identified Be
stars in NGC 2439 via our 2-CD and 1 previously undetected 
candidate Be star.  Spectral types for two of the known Be stars are given in
\citet{sle85}: star White $\# 6$ \citep{whi75} is a B2 V, which generally 
agrees with
our rough B0 classification of this object (NGC 2439:WBBe 1 in Table \ref{photdetails}), and star
White $\# 81$ \citep{whi75} is a B1 V, which generally agrees with our rough 
B0 classification of this star (NGC 2439:WBBe 2 in Table \ref{photdetails}).  

\subsection{ELHC Fields}

As previously described, we have detected ELHC \# 1,3,4,6,7,8,12,13, and 19 
as candidate emission-line objects via 2-CDs and failed to detect ELHC 5, 11, and 20.
This 75\% detection rate offers supporting evidence that ELHC stars might be bonafide
pre-main-sequence objects.  Recent low-resolution spectroscopic and JHK photometric 
observations of some of these ELHCs \citep{dew05} have revealed that some objects might have nearby, 
previously unresolved neighbors; furthermore, the lack of a strong near IR excess and lack of
forbidden emission lines has led these authors to question whether many of these ELHCs
are truly Herbig Ae/Be stars or whether they are classical Be stars.

With respect to our results, we also had noted that ELHC 11 seemed to be comprised
of 2 components.  The low-resolution spectroscopy of ELHC 11 and 20, which were
both not detected as emission-line objects in our study, revealed these sources to
have H$\alpha$ absorption lines slightly filled in with emission \citep{dew05}.  Because
the 2-CD technique fails to detect weak H$\alpha$ emitters, it is not surprising that
we failed to detect these objects.  Based on these results, we would predict that
when spectroscopic observations of ELHC 5 are made, they will also reveal a partially
filled in absorption line at H$\alpha$.  In a future publication, we will present the results of
moderate resolution spectroscopic and near-IR photometric observations of many of these 
ELHC stars which were obtained to further probe the true nature of these objects.

\subsection{Limitations of the 2-CD technique}

We briefly discussed some of the known limitations of the 2-CD technique in the 
Introduction, and now consider this situation in more detail.  The 2-CD 
technique typically 
assumes that all stars within an appropriate range of colors that have a
(R-H$\alpha$) magnitude in excess of some threshold are ``Be stars''.  
However, as many Galactic O, B, and A type supergiants are known to show 
frequent H$\alpha$ emission, one expects that the detection of such objects via the 2-CD technique would
artificially inflate the number of detected ``Be stars''.  For example,
\citet{ols01} obtained follow-up spectroscopy of star S132 in the LMC cluster
LH 72, which they identified as a Be star from a 2-CD, and classified this
object as a B8 Ia star.  The luminosity class of S132 excludes it, by definition, from being a 
classical Be star.  Similarly, 
\citet{kel99} identified NGC 346:KWBBe 13 as a ``Be star'' via a 2-color diagram. 
  Unfortunately this object is the well known HD 5980 Wolf-Rayet/Luminous 
Blue Variable system, thus it is abundantly clear that it has been mis-classified by the simple 2-CD method
 as a (classical) ``Be star''.  In principle, one would expect that the location of
these objects on a CMD would identify them as ``contaminants''; however, this is
clearly not always the case.  It is possible that these stars are background objects with respect to 
the clusters with which they have been associated.  It is also possible that a significant amount 
of their radiation is attenuated by a localized region of dust.  These two examples demonstrate that it is not  
unreasonable to expect OBA supergiants to ``contaminate'' 2-CD detections.
We suggest the frequency of contamination by such objects is likely to be greater when 
one uses the 2-CD technique to investigate field star populations \citep{kel99, kel01}, compared to cluster populations, 
as the evolutionary age of a cluster population should restrict the range of spectral types 
of OBA supergiants present.  

Recent follow-up low-resolution spectroscopic and near IR photometric observations of 
some ELHCs \citep{dew05} have raised issues with their classification as Herbig Ae/Be objects.
Even assuming the most pessimistic interpretation of the results of \citet{dew05}, i.e.  
that ELHC 7 is the only true Herbig Ae/Be star of these candidates as it is the only one
which exhibits a large near IR excess, the implications of these results regarding the proper interpretation 
of 2-CDs can not be ignored.  We detected ELHC 7 on a 2-CD as being an excess H$\alpha$ 
emitter; therefore, we have solid evidence that Herbig Ae/Be stars may be detected via
this technique.  Clearly, it is not sufficient to merely label all excess H$\alpha$ emitters
detected via 2-CDs as ``Be stars'' as is nearly universally done in the literature.  
Given a cluster of sufficient age, and assuming coeval development, one would not  
expect to find any remaining pre-main-sequence stars;  
hence, this type of contamination may not be an issue.  However, when examining
the background field Be population surrounding a cluster using the 2-CD
technique \citep[see][]{kel99}, we assert that one must be wary of
detecting Herbig Ae/Be stars, which would serve to artificially inflate
the ratio of detected classical Be stars.  

\citet{kel00}, \citet{ols01}, 
Olsen (2003, private communication), and the present study all found evidence
that diffuse background contamination likely inflates the number of true
classical Be stars detected via the 2-CD technique.  A typical form of these spurious detections
we encountered were random, quasi point-source-like pockets of emission which 
were misconstrued as stars, especially when the \textit{daofind} image roundness 
parameters were slightly relaxed.  One also might consider how diffuse sky nebulosity
affects the measurement of real sources.  The IRAF photometry methods we used
sampled the sky background within an annulus of a given radius around sources and
used the median pixel value as an estimate of the background.  In extremely
variable backgrounds, for example containing H II filamentary structures, it is not
always clear that this technique fully accounts for the true background present.
While we do not propose a better technique to estimate background fluxes in regions
of complex nebulosity, we feel it is important to 
remind the reader of some of the difficulties associated with observations of embedded objects
 when considering the results of these and similar 2-CD investigations.  The addition of other 
independent techniques, such as IR photometry and broadband or spectropolarimetry, could 
provide much needed cross-checks on such identifications.

\section{Summary}

We have presented B, V, R, and H$\alpha$ photometry of 5 LMC, 8 LMC, and 3 
Galactic clusters. 
Plotting these data on 2 color diagrams, we identified the fractional content of
candidate Be stars in each cluster, i.e the ratio of Be/(B+Be).  
From these data we found:

1) Candidate Be stars appear in significant numbers in all clusters studied.
We provide basic photometric information and astrometric RA and Dec 
coordinates for these candidate Be stars to facilitate follow-up investigations.

2) Four clusters with ages less than 10 Myr were found to have significant
numbers of candidate Be stars, with crude spectral types ranging from B0 to B5.  
The fractional Be content of these young clusters was similar to the nominal value 
found in our Galaxy, $\sim$17\%.  Clearly, these results could dramatically alter our current understanding
of the role evolutionary age plays in the development of the Be phenomenon 
\textit{if} we can show that these objects are truly classical Be stars.

3) The fractional content of early-type candidate Be stars appears to be 
significantly enhanced in clusters of age 10-25 Myr old.  Clusters older than
25 Myr also appear to have enhanced levels of early- and later-type candidate
Be stars.  We suggest that the spin-up of stars as they evolve along the 
main sequence and/or the enhanced main sequence lifetime of rapidly rotating
stars compared to slow rotators may explain this effect.

4) From inspection of all B0 to B3 type candidate Be stars in clusters with 
ages 10-25 Myr, we find evidence that the Be phenomenon is more prevalent 
in low metallicity environments.  The additional
statistics present in this study lowers the average fractional Be content of SMC
clusters from 39\% \citep{mae99} to 32\%.  

5) We examined the Be population in two clusters, LH 72 and NGC 2439, which
had their Be population previously investigated, and found evidence of
previously unreported candidate Be stars.  The variable nature of Be stars likely accounts 
for some of these differences, although spurious detections may play a more dominant role
in explaining the differences observed in LH 72.

6) 75\% of previously suggested ELHCs were detected as candidate emission-line objects via
our 2-color diagrams.  These detections: a) offer 
evidence that these objects do emit excess H$\alpha$ emission and hence
offer supporting evidence that they might be pre-main-sequence objects; and b) illustrate that
the 2 color diagram technique will identify other types of emission-line
objects, besides classical Be stars.

7) We suggest that all objects identified by our 2 color diagrams, as well
as similarly identified objects in the literature, should be classified as ``candidate Be stars''.   
Herbig Ae/Be, B[e], and OBA supergiants, which are of a fundamentally different
nature than classical Be stars, often exhibit H$\alpha$ emission and thus
also may be detected via the 2-CD technique.  Complementary techniques, such as polarimetry, IR photometry, 
and optical/IR spectroscopy can help to remove some of this potential confusion.

\acknowledgments

We thank the referee, Douglas Gies, for comments which helped to improve this paper.
We also thank the NOAO TAC for awarding observing time for this project.  JPW
acknowledges support from a NASA GSRP fellowship (NGT5-50469) and thanks NOAO for 
supporting his travel to CTIO.  KSB is a Cottrell Scholar of the Research
Corporation and gratefully acknowledges their support.  This work has been
supported in part by NASA LTSA grant NAG5-8054 to the University of Toledo.
This research has made use of the SIMBAD database operated at CDS, Strasbourg,
France, and the NASA ADS system.

\clearpage
\newpage

\figsetstart
\figsetnum{1}
\figsettitle{2 Color Diagrams (2-CDs)}

\figsetgrpstart
\figsetgrpnum{1.1}
\figsetgrptitle{Bruck 60}
\figsetplot{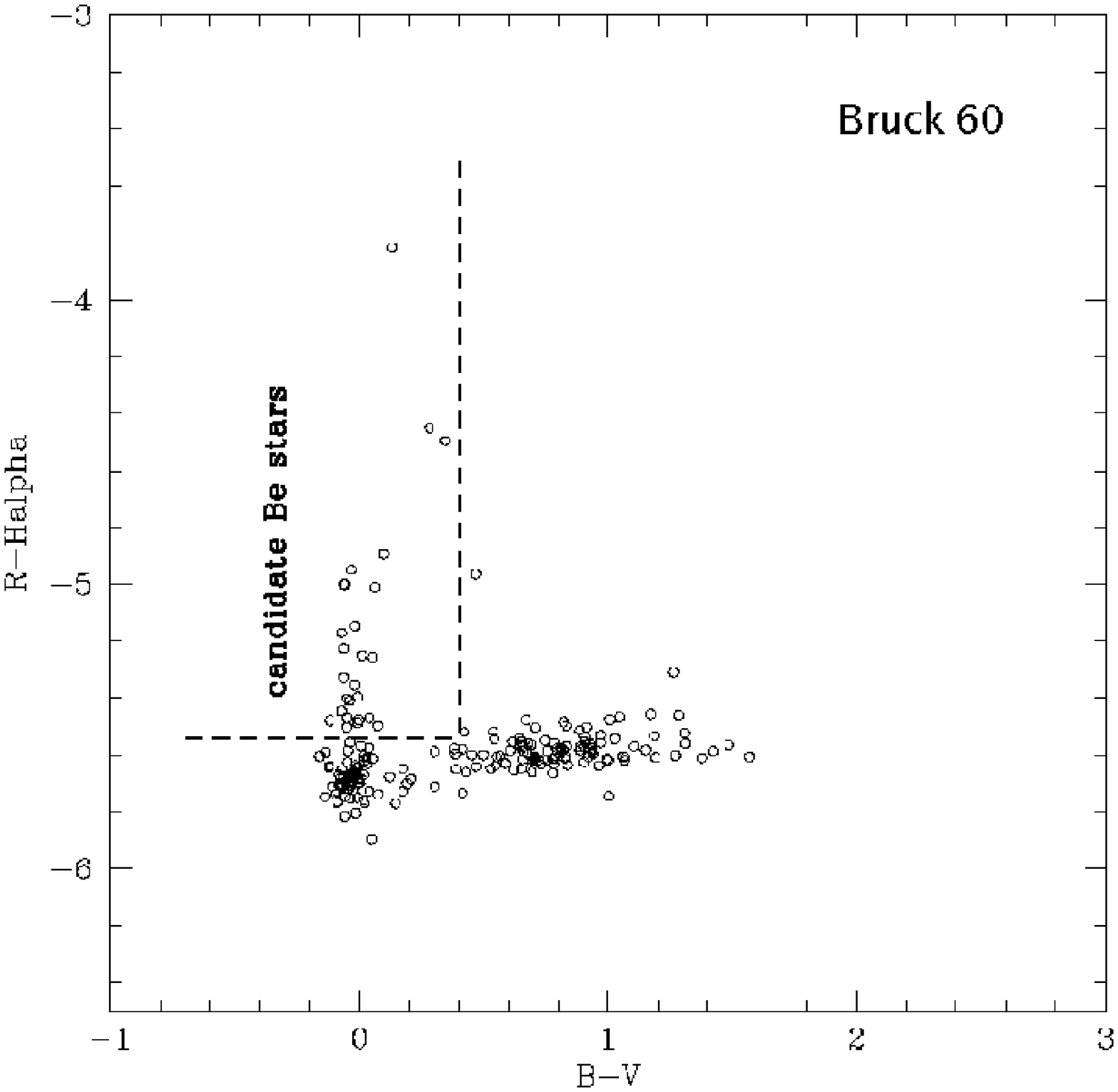}
\figsetgrpnote{2 Color Diagrams}
\figsetgrpend

\figsetend

\figsetstart
\figsetnum{2}
\figsettitle{Color Magnitude Diagrams}

\figsetgrpstart
\figsetgrpnum{2.1}
\figsetgrptitle{Bruck 60}
\figsetplot{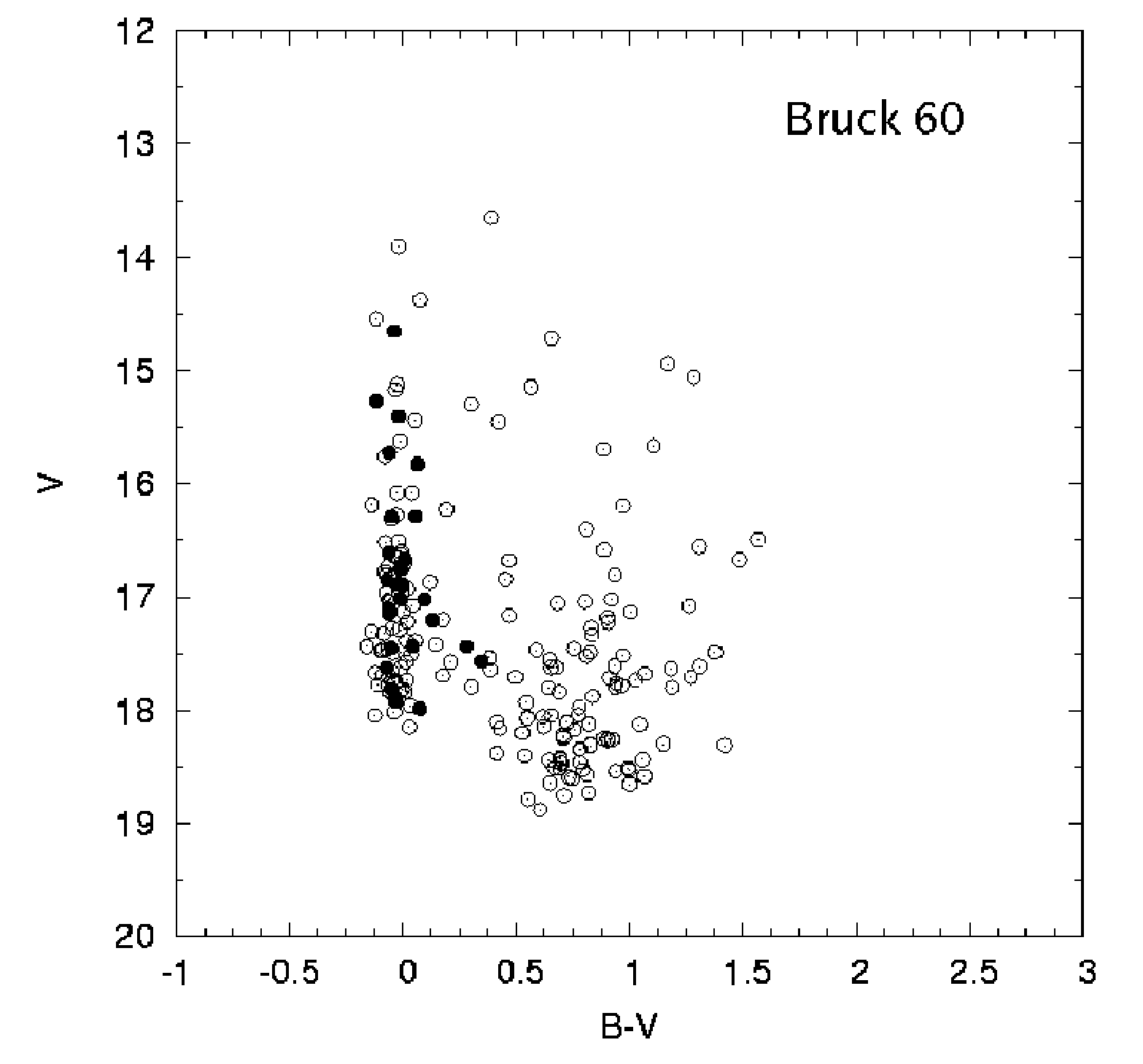}
\figsetgrpnote{Color Magnitude Diagrams}
\figsetgrpend

\figsetend

\clearpage

\clearpage

\clearpage
\begin{table}
\caption{Basic Properties of the Observations}
\tiny
\begin{tabular}{lccllcc}
\tableline
Cluster Name & Location & Date & Filter  & Exposure Times & log Cluster Age & 
E(B-V) \\ 
\tableline
ELHC-2$^{3}$ & LMC & 2002 October 16 & B,V,R,H$\alpha$ & 60,15,15,120 &
\nodata & \nodata \\
\nodata & \nodata & \nodata & B,V,R,H$\alpha$,H$\alpha$ & 300,100,100,600,600 &
\nodata & \nodata \\
\nodata & \nodata & \nodata & B,V,R,H$\alpha$,H$\alpha$ & 300,300,300,600,600 &
\nodata & \nodata \\
ELHC-3$^{4}$ & LMC & 2002 October 16 & B,V,R,H$\alpha$ & 60,15,15,120 & 
\nodata & \nodata \\
\nodata & \nodata & \nodata & B,V,R,H$\alpha$,H$\alpha$ & 300,100,100,600,600
 & \nodata & \nodata \\
LH 72 & LMC & 2002 October 12 & B,V,R,H$\alpha$ & 60,15,15,120 &
6.7-7.2$^{5}$ & 0.09$^{5}$ \\
\nodata & \nodata & \nodata & B,V,R,H$\alpha$,H$\alpha$ & 300,100,100,600,600 &
\nodata & \nodata \\
\nodata & \nodata & \nodata & B,V,R,H$\alpha$,H$\alpha$ & 600,600,600,600 & 
\nodata & \nodata \\
NGC 1850 & LMC & 2002 October 13 & B,V,R,H$\alpha$ & 60,10,10,15 
& 7.7$^{1,7}$-7.8$^{1,6}$, 6.6$^{2,7}$-6.9$^{2,6}$ & 0.18$^{6}$ \\
\nodata & \nodata & \nodata & B,V,R,H$\alpha$ & 300,100,100,120 & \nodata &
\nodata \\
\nodata & \nodata & \nodata & B,V,R,H$\alpha$,H$\alpha$ & 600,600,600,600,600 &
\nodata & \nodata \\
NGC 1858 & LMC & 2002 October 16 & B,V,R,H$\alpha$ & 60,15,15,120  
& 6.9$^{6}$ & 0.15$^{6}$ \\
\nodata & \nodata & \nodata & B,V,R,H$\alpha$,H$\alpha$ & 300,100,100,600,600 & 
\nodata & \nodata \\
\nodata & \nodata & \nodata & B,V,R,H$\alpha$,H$\alpha$ & 600,100,600,600,600 &
\nodata & \nodata \\
NGC 1955 & LMC & 2002 October 14 & B,V,R,H$\alpha$ & 60,15,15,120  
& 6.82$^{8}$ & 0.09$^{8}$ \\
\nodata & \nodata & \nodata & B,V,R,H$\alpha$,H$\alpha$ & 600,300,300,600,600 
& \nodata & \nodata \\
NGC 2027 & LMC & 2002 October 15 & B,V,R,H$\alpha$ & 60,15,15,15 
& 7.06$^{8}$ & 0.05$^{8}$ \\
\nodata & \nodata & \nodata & B,V,R,H$\alpha$ & 300,100,100,120 & \nodata &
 \nodata \\
\nodata & \nodata & \nodata & B,V,R,H$\alpha$,H$\alpha$ & 600,600,600,600,600
& \nodata & \nodata \\
NGC 2186 & MWG & 2002 October 14 & B,V,R,H$\alpha$ & 30,15,15,20 & 
7.738$^{13}$ & 0.31$^{15}$ \\
\nodata & \nodata & \nodata & B,V,R,H$\alpha$ & 200,100,100,120 & \nodata & 
\nodata \\
NGC 2383 & MWG & 2002 October 12 & B,V,R,H$\alpha$ & 5,5,5,10 & 7.4$^{13}$,
 8.6$^{14}$ & 0.22$^{14}$ \\
\nodata & \nodata & \nodata & B,V,R,H$\alpha$ & 30,30,30,60 & \nodata & 
\nodata \\
\nodata & \nodata & \nodata & B,V,R,H$\alpha$ & 120,120,120,240 & \nodata &
\nodata \\
NGC 2439 & MWG & 2002 October 12 & B,V,R,H$\alpha$ & 3,5,3,3 & 7.3$^{12}$, 
7.82$^{13}$ & 0.37$^{12}$ \\
\nodata & \nodata & \nodata & B,V,R,H$\alpha$ & 20,20,20,20 & \nodata & 
\nodata \\
\nodata & \nodata & \nodata & B,V,R,H$\alpha$ & 120,120,120,120 & \nodata & 
\nodata \\
Bruck 60 & SMC & 2002 October 15 & B,V,R,H$\alpha$ & 60,15,15,120 & 
 7.8$^{11}$ & $0.037^{18}$ \\
\nodata & \nodata & \nodata & B,V,R,H$\alpha$,H$\alpha$ & 300,100,100,600,600
& \nodata & \nodata \\
\nodata & \nodata & \nodata & B,V,R,H$\alpha$,H$\alpha$ & 600,600,600,600,600
& \nodata & \nodata \\
Bruck 107 & SMC & 2002 October 15 & B,V,R,H$\alpha$ & 60,15,15,120 
& 8.1$^{11}$ & $0.037^{18}$ \\
\nodata & \nodata & \nodata & B,V,R,H$\alpha$,H$\alpha$ & 300,600,100,600,600
& \nodata & \nodata \\
\nodata & \nodata & \nodata & B,V,R,H$\alpha$,H$\alpha$ & 600,600,600,600,600
& \nodata & \nodata \\
HW 43 & SMC & 2002 October 13 & B,V,R,H$\alpha$ & 60,15,15,120 
& 7.9$^{11}$ & $0.037^{18}$ \\
\nodata & \nodata & \nodata & B,V,R,H$\alpha$,H$\alpha$ & 300,100,100,600,600
& \nodata  & \nodata \\
\nodata & \nodata & \nodata & B,V,R,H$\alpha$,H$\alpha$ & 600,600,600,600,600
& \nodata & \nodata \\ 
NGC 371 & SMC & 2002 October 12 & B,V,R,H$\alpha$ & 60,15,15,120 
& 6.74 $^{9}$ & 0.08$^{9}$ \\
\nodata & \nodata & \nodata & B,V,R,H$\alpha$,H$\alpha$ & 300,100,100,600,600
& \nodata & \nodata \\
\nodata & \nodata & \nodata & B,V,R,H$\alpha$,H$\alpha$ & 600,600,600,600,600
& \nodata & \nodata \\
NGC 456 & SMC & 2002 October 14 & B,V,R,H$\alpha$ & 60,15,15,120 
& 7.0 $^{11}$ & 0.27$^{16}$ \\
\nodata & \nodata & \nodata & B,V,R,H$\alpha$,H$\alpha$ & 300,100,100,600,600
& \nodata & \nodata \\
\nodata & \nodata & \nodata & B,V,R,H$\alpha$,H$\alpha$ & 600,600,600,600,600
& \nodata & \nodata \\
NGC 458 & SMC & 2002 October 13 & B,V,R,H$\alpha$,H$\alpha$ & 60,15,15,600,600
& 7.7$^{11}$, 8.0-8.18$^{10}$ & 0.04$^{17}$ \\
\nodata & \nodata & \nodata & B,V,R,H$\alpha$,H$\alpha$ & 300,100,100,600,600
& \nodata & \nodata \\
\nodata & \nodata & \nodata & B,V,R,H$\alpha$,H$\alpha$ & 600,600,600,600,600
& \nodata & \nodata \\
NGC 460 & SMC & 2002 October 16 & B,V,R,H$\alpha$ & 60,15,15,120  
& 7.3$^{11}$ & 0.12$^{16}$ \\
\nodata & \nodata & \nodata & B,V,R,H$\alpha$,H$\alpha$ & 300,100,100,600,600
& \nodata & \nodata \\
\nodata & \nodata & \nodata & B,V,R,H$\alpha$,H$\alpha$ & 600,600,600,600,600
& \nodata & \nodata \\
NGC 465 & SMC & 2002 October 16 & B,V,R,H$\alpha$ & 60,15,15,120  
& \nodata & 0.09$^{16}$ \\
\nodata & \nodata & \nodata & B,V,R,H$\alpha$,H$\alpha$ & 300,100,100,600,600
& \nodata & \nodata \\
\nodata & \nodata & \nodata & B,V,R,H$\alpha$,H$\alpha$ & 600,600,600,600,600
& \nodata & \nodata \\
\tableline
\tablecomments{Summary of our CTIO 0.9 m photometric observations.  Note that
$^{1}$ refers to NGC 1850A and $^{2}$ refers to NGC 1850B.  $^{3}$ and 
$^{4}$ denote observations of LMC fields which were reported to contain
candidate HAeBe stars\citep{lam99,dew02}.  $^{18}$ denote clusters for which
no E(B-V) value was found in the literature, thus we assign the mean reddening
value for the SMC determined by \citet{sch98} to these clusters.  
References cited are:
$^{5}$ \citet{ols01}, $^{6}$ \citet{val94}, $^{7}$ \citet{gil94},
$^{8}$ \citet{dol98}, $^{9}$ \citet{mas00}, $^{10}$ \citet{mat02},
$^{11}$ \citet{hod83},
$^{12}$ \citet{whi75}, $^{13}$ \citet{lyn95}, $^{14}$ \citet{sub99},
$^{15}$ \citet{mof75}, $^{16}$ \citet{hi94b}, $^{17}$ \citet{alc03}}
\label{allsum}
\end{tabular}
\end{table}

\clearpage
\begin{table}
\caption{Coefficients of Standard Star Transformations}
\scriptsize
\begin{tabular}{cccc}
\tableline
Date & B$_{0}$ & B$_{1}$ & B$_{2}$ \\
\tableline
2002 October 12 & 2.537 $\pm 0.016$ & 0.30 & 0.118 $\pm 0.019$ \\
2002 October 13 & 2.246 $\pm 0.015$ & 0.30 & 0.109 $\pm 0.019$ \\
2002 October 14 & 2.258 $\pm 0.014$ & 0.30 & 0.132 $\pm 0.017$ \\
2002 October 15 & 2.477 $\pm 0.041$ & 0.30 & 0.076 $\pm 0.052$ \\
2002 October 16 & 2.231 $\pm 0.007$ & 0.30 & 0.093 $\pm 0.009$ \\
\tableline
\nodata & V$_{0}$ & V$_{1}$ & V$_{2}$ \\
\tableline
2002 October 12 & 2.467 $\pm 0.012$ & 0.15 & -0.010 $\pm 0.014$ \\
2002 October 13 & 2.101 $\pm 0.012$ & 0.15 & -0.017 $\pm 0.015$ \\
2002 October 14 & 2.119 $\pm 0.008$ & 0.15 & -0.023 $\pm 0.010$ \\
2002 October 15 & 2.324 $\pm 0.030$ & 0.15 & -0.295 $\pm 0.049$ \\
2002 October 16 & 2.086 $\pm 0.006$ & 0.15 & -0.018 $\pm 0.007$ \\
\tableline
\nodata & R$_{0}$ & R$_{1}$ & R$_{2}$ \\
\tableline
2002 October 12 & 2.495 $\pm 0.019$ & 0.08 & 0.077 $\pm 0.040$ \\
2002 October 13 & 2.206 $\pm 0.011$ & 0.08 & -0.001 $\pm 0.021$ \\
2002 October 14 & 2.216 $\pm 0.006$ & 0.08 & -0.016 $\pm 0.013$ \\
2002 October 15 & 2.224 $\pm 0.020$ & 0.08 & -0.013 $\pm 0.043$ \\
2002 October 16 & 2.206 $\pm 0.005$ & 0.08 & -0.015 $\pm 0.011$ \\
\tableline

\tablecomments{Photometric transformation coefficients.}
\label{trans}
\end{tabular}
\end{table}

\clearpage
\begin{table}
\caption{Summary of Detected Candidate Be Stars}
\scriptsize
\begin{tabular}{p{1.5cm} p{.8cm} p{.8cm} p{1.3cm} p{1.4cm} p{1.4cm} p{1.5cm} p{0.8cm} p{.9cm} p{.8cm}}
\tableline
Cluster &  V$<$14 & V14 & V15 & V16 & V17 & V18 & V$\geq$19 & \# Be & \# MS \\
\tableline
Bruck 60 & 0/1 & 1/3 & 4/9 & 7(8)/29 & 9(13)/53 & 0/3 & 0/0 & 21(26) & 98 \\
Bruck 107$^{1}$ & 0/0 & 1/1 & 1/1 & 1/5 & 5/11 & 4/21 & 0/2 & 12 & 41 \\
Bruck 107$^{2}$ & 0/0 & 1/2 & 0/1 & 2/5 & 0/6 & 3(4)/19 & 0/2 & 6(7) & 35 \\
HW 43 & 0/0 & 0/0 & 0/0 & 0/1 & 1/11 & 4/31 & 2/9 & 7 & 52 \\
LH 72 & 0/4 & 5(6)/22 & 7/40 & 5(9)/63 & 15(17)/86 & 7(11)/71 & 0/0 & 39(50)
 & 286 \\
NGC 371 & 1/4 & 9/52 & 11(12)/124 & 33(35)/250 & 43(47)/432 & 21(25)/222 & 0/0
 & 118(129) & 1084 \\
NGC 456 & 0/0 & 3/4 & 1/5 & 3/16 & 12(13)/32 & 1/28 & 2/5 & 22(23) & 90 \\
NGC 458 & 0/0 & 0/0 & 0/4 & 2/4 & 4(6)/25 & 19/58 & 3/7 & 28(30) & 98 \\
NGC 460 & 0/0 & 1(2)/7 & 3(5)/16 & 2(3)/19 & 6(8)/36 & 2(3)/17 & 0/0 & 14(21)
 & 95 \\ 
NGC 465 & 0/0 & 1/14 & 3/22 & 1/32 & 4/75 & 1/65 & 1/7 & 11 & 215 \\
NGC 1850 & 0/0 & 1/45 & 5/41 & 19(20)/134 & 44(49)/199 & 17/72 & 0/0 & 
86(92) & 492 \\
NGC 1858 & 0/1 & 1/13 & 5/21 & 12(15)/37 & 16(17)/35 & 1/2 & 0/0 & 35(39)
 & 109 \\
NGC 1955 & 0/3 & 0/10 & 1/23 & 5/44 & 13(15)/46 & 2(3)/15 & 0/0 & 21(24)
 & 141 \\
NGC 2027 & 3/10 & 4/51 & 10/78 & 11/143 & 11/245 & 4(7)/324 & 0/0 & 43(46) 
& 851 \\
ELHC 2$^{3}$ & 0/11$^{5}$ & 3/37$^{5}$ & 13/77$^{5}$ & 36/341$^{5}$ & 
78/1013$^{5}$ & 53/771$^{5}$ & 0/0$^{5}$ & 183$^{5}$ & 2250$^{5}$ \\
ELHC 3$^{4}$ & 1/14$^{5}$ & 7/67$^{5}$ & 24/151$^{5}$ & 35/613$^{5}$ & 
55/1654$^{5}$ & 31/1137$^{5}$ & 0/0$^{5}$ & 153$^{5}$ & 3636$^{5}$ \\
NGC 2186 & 5/9 & 0/0 & 0/0 & 0/0 & 0/0 & 0/0 & 0/0 & 5 & 9 \\
NGC 2383 & 3/14 & 0/26 & 0/48 & 0/18 & 0/0 & 0/0 & 0/0 & 3 & \nodata$^{6}$ \\
NGC 2439 & 4/73 & 1/58 & 0/0 & 0/0 & 0/0 & 0/0 & 0/0 & 5 & 131 \\ 
\tableline
\tablecomments{A summary of candidate Be stars found in this study, grouped in bins
of observed V magnitude.  Data in the V15 column, for example, correspond to 
candidate Be stars having 15.0 $\leq$ V $\leq$ 15.9.  Detection 
summaries listed in parenthesis are ``possible detections'' 
and should be viewed as less reliable than detection summaries without parenthesis.
$^{1}$ assumes a cluster diameter of 3.5 arc-minutes.  $^{2}$ is based upon a ring
located at a radius of 1$^{'}$.75 to 2$^{'}$.88 from the cluster center,
 which \citet{kon80}
claim represents the background population.  $^{3}$ and $^{4}$ represent
observations of LMC fields which \citet{lam99} and \citet{dew02} claim contain numerous
candidate Herbig Ae/Be stars.  $^{5}$ indicates the detection of ``candidate
emission-line stars'' and is not meant to reflect the field's
classical Be star content.   $^{6}$ The number of B-type main sequence stars in NGC 2383 is 
not calculated, owing to the large uncertainty in the distance to this cluster.  }
\label{photsum}
\end{tabular}
\end{table}

\clearpage
\begin{table}
\caption{Photometric Properties of the Candidate Be Stars}
\scriptsize
\begin{tabular}{cccccccccc}
\tableline
Name & RA (2000) & Dec (2000) & V & V$_{err}$ & (B-V) & (B-V)$_{err}$ &
 (R-H$\alpha$) &  (R-H$\alpha$)$_{err}$ & Status \\
\tableline
Bruck 60:WBBe 1 & 0 51 44.0 & -73 14 27.7 & 16.88 & 0.02 &
 -0.02 & 0.02 & -5.35 & 0.04 & C \\
Bruck 60:WBBe 2  & 0 51 58.8 & -73 15  3.1 & 16.67  & 0.01  &  0.01 &  0.01 &
 -5.25 & 0.03 & C \\
Bruck 60:WBBe 3 &   0 51 35.4 & -73 12 42.1 & 16.29 &  0.01  & -0.05 &  0.01 &
 -5.47 & 0.02 & C \\
Bruck 60:WBBe 4  &  0 51 31.8  & -73 13  9.2   & 17.02  & 0.02  & -0.01 &  0.02 &
 -5.49 & 0.05 & C \\
Bruck 60:WBBe 5   & 0 51 42.8 & -73 13 27.4 & 15.83  & 0.01 &   0.07 &  0.01 & 
-5.01 & 0.02 & C \\
\tableline
\tablecomments{A full version of this table will appear in the online Journal.  The ``Status'' column 
designates the Be star classification we have assigned each object, ``C'' = candidate Be stars and 
``P'' = possible candidate Be star.}
\label{photdetails}
\end{tabular}
\end{table}

\clearpage
\begin{table}
\caption{Spectral Types of Candidate Be Stars in Every Cluster Environment}
\scriptsize
\begin{tabular}{p{1.35cm} p{0.4cm} p{1.75cm} p{1.75cm} p{1.75cm} p{1.65cm} p{1.75cm} p{1.75cm}}
\tableline
Cluster &  Age & B0 & B1 & B2 & B3 & B4 & B5 \\
\tableline
NGC 371 & vy & 19/158 (12\%) & 16/136 (12\%) & 35/265 (13\%) & 16/177 (9\%)
 & 29/202 (14\%) & 14/146 (10\%) \\
LH 72 & vy & 11/48 (23\%) & 3/32 (9\%) & 9/50 (18\%) & 5/28 (18\%) &
8/41 (20\%) & 7/34 (21\%) \\
NGC 1858 & vy & 5/34 (15\%) & 6/18 (33\%) & 14/30 (47\%) & 8/13 (62\%) &
 5/12 (42\%) & 1/2 (50\%) \\
NGC 1955 & vy & 0/23 (0\%) & 2/24 (8\%) & 4/33 (12\%) & 3/16 (19\%)&
 9/23 (39\%) & 4/18 (22\%) \\
NGC 456 & y & 6/14 (43\%) & 2/13 (15\%) & 7/19 (37\%) & 5/12 (42\%) &
 1/18 (6\%) & 0/10 (0\%) \\
NGC 460 & y & 7/23 (30\%) & 2/11 (18\%) & 4/26 (15\%) & 3/13 (23\%) & 
5/17 (29\%) & 0/3 (0\%) \\
NGC 2027 & y & 13/92 (14\%) & 4/57 (7\%) & 9/110 (8\%) & 6/70 (9\%) & 
5/116 (4\%) & 2/103 (2\%) \\
Bruck 60 & o & 4/11 (36\%) & 3/10 (30\%) & 10/30 (30\%) & 5/22 (23\%) &
 4/24 (17\%) & 0/1 (0\%) \\
Bruck 107 & o & 2/2 (100\%) & 1/4 (25\%) & 0/3 (0\%) & 2/3 (67\%) & 
3/6 (50\%) & 0/5 (0\%) \\
HW 43 & o & 0/0 (0\%) & 0/0 (0\%) & 0/3 (0\%) & 1/2 (50\%) & 1/9 (11\%) 
& 2/6 (30\%) \\
NGC 458 & o & 0/1 (0\%) & 0/3 (0\%) & 2/6 (30\%) & 3/11 (27\%) & 3/15 (20\%) 
& 7/24 (29\%) \\
NGC 1850 & o? & 4/82 (5\%) & 7/57 (12\%) & 28/139 (20\%) & 17/72 (24\%) 
& 25/100 (25\%) & 11/42 (26\%) \\
NGC 465 & \nodata & 4/33 (12\%) & 1/22 (5\%) & 1/34 (3\%) & 1/27 (4\%) & 
3/43 (7\%) & 0/21 (0\%) \\
\tableline
\tablecomments{Spectral types were assigned using the calibrations of
\citet{zor97}, following the methods outlined by \citet{gre97}: 
$B_{0}: M_{V} < -3.25$, $B_{1}:  -3.25<M_{V}<-2.55$, $B_{2}:  
-2.55<M_{V}<-1.8$, $B_{3}: -1.8<M_{V}<-1.4$, $B_{4}: -1.4<M_{V}<-0.95$, 
$B_{5}: -0.95<M_{V}<-0.6$.  \citet{gre97} offers an excellent discussion
of the numerous uncertainties inherent in assigning spectral types via this
technique.  In column 2, we group clusters into 3 age groups: very young (vy)
: $6.7 < log (t) < 6.9$; young (y): $7.0 < log(t) < 7.4$; and old (o):
$7.5 < log(t) < 8.2$.  As discussed in Section 3.2.3, NGC 1850 likely
contains populations of multiple epochs, thus we assign it a designation of
``old'' with caution.}
\label{photspectypesum}
\end{tabular}
\end{table}

\clearpage
\begin{table}
\caption{Average Number of Early- and Later-Type Candidate Be Stars in Every Cluster}
\scriptsize
\begin{tabular}{ccccc}
\tableline
Cluster &  Age & Location & Early-Type & Later-Type \\
\tableline
NGC 371 & vy & SMC & 86/736 (12\% $\pm$ 1\%) & 43/348 (12\% $\pm$ 2\%) \\
NGC 346$^{2}$ & vy & SMC & 11/78 (14\% $\pm$ 4\%) & \nodata \\
LH 72 & vy & LMC & 28/158 (18\% $\pm$ 3\%) & 15/75 (20\% $\pm$ 5\%) \\
NGC 1858 & vy & LMC & 33/95 (35\% $\pm$ 5\%) & \nodata \\
NGC 1955 & vy &LMC & 9/96 (9\% $\pm$ 3\%) & 13/41 (32\% $\pm$ 7\%) \\
NGC 330$^{2}$ & y & SMC & 27/79 (34\% $\pm$ 5\%) & \nodata \\
NGC 330$^{1}$ & y & SMC & 50/128 (39\% $\pm$ 4\%) & \nodata \\
NGC 456 & y & SMC & 20/58 (34\% $\pm$ 6\%) & 1/28 (4\% $\pm$ 4\%) \\
NGC 460 & y & SMC & 16/73 (22\% $\pm$ 5\%) & 5/20 (25\% $\pm$ 10\%) \\
NGC 2006$^{1}$ & y & LMC & 10/35 (29\% $\pm$ 8\%) & \nodata \\
NGC 2004$^{1}$ & y & LMC &  25/130 (19\% $\pm$ 3\%) & \nodata \\
NGC 2027 & y & LMC & 32/329 (10\% $\pm$ 2\%) & 7/219 (3\% $\pm$ 1\%) \\
Hodge 301$^{1}$ & y & LMC & 10/44 (23\% $\pm$ 6\%) & \nodata \\
NGC 1818A$^{1}$ & y & LMC & 34/94 (36\% $\pm$ 5\%) & \nodata \\
NGC 1948$^{1}$ & y & LMC & 11/101 (11\% $\pm$ 3\%) & \nodata \\
NGC 2100$^{1}$ & y & LMC & 19/67 (28\% $\pm$ 6\%) & \nodata \\
NGC 1818$^{2}$ & y & LMC & 19/92 (21\% $\pm$ 4\%) & \nodata \\
NGC 2004$^{2}$ & y & LMC & 16/124 (13\% $\pm$ 3\%) & \nodata \\
SL 538$^{1}$ & y & LMC & 11/46 (24\% $\pm$ 6\%) & \nodata \\
NGC 457$^{1}$ & y & MW ext. & 4/28 (14\% $\pm$ 6\%) & \nodata \\
NGC 663$^{1}$ & y & MW ext. & 12/35 (34\% $\pm$ 8\%) & \nodata \\
NGC 869$^{1}$ & y & MW ext. & 3/42 (7\% $\pm$ 4\%) & \nodata \\
NGC 884$^{1}$ & y & MW ext. & 6/28 (21\% $\pm$ 8\%) & \nodata \\
NGC 2439$^{1}$ & y & MW ext. & 5/31 (16\% $\pm$ 6\%) & \nodata \\
NGC 3293$^{1}$ & y & MW int. & 1/37 (3\% $\pm$ 3\%) & \nodata \\
NGC 3766$^{1}$ & y & MW int. & 10/42 (24\% $\pm$ 7\%) & \nodata \\
NGC 4755$^{1}$ & y & MW int. & 3/47 (6\% $\pm$ 3\%) & \nodata \\
Bruck 60 & o & SMC & 22/73 (30\% $\pm$  5\%) & 4/25 (16\% $\pm$ 7\%) \\
NGC 458 & o & SMC & 5/21 (24\% $\pm$ 10\%) & 10/39 (26\% $\pm$ 7\%) \\
NGC 1850 & o? & LMC & 56/350 (16\% $\pm$ 2\%) & 36/142 (25\% $\pm$ 4\%) \\
\tableline
\tablecomments{The superscript ``1'' denotes data culled from \citet{mae99} while
the superscript ``2'' denotes data culled from \citet{kel99}.  All of the cluster data 
taken from the literature correspond to ``young'' clusters \citep{mae99}, although we
note that age estimates differing by up to a factor of 2 from those listed in \citet{mae99} 
may be found elsewhere in the literature.}
\label{earlylate}
\end{tabular}
\end{table}

\clearpage
\begin{table}
\caption{Number of Early- and Later-Type Candidate Be Stars Averaged by Common Metallicity Environment}
\begin{tabular}{cccccccc}
\tableline
Region &  Age & Early-Type & Size & Later-Type & Size \\
\tableline
SMC & vy & 13\% $\pm$ 3\% & 2 & 12\% $\pm$ 2\% & 1 \\
LMC & vy & 20\% $\pm$ 4\% & 3 & 26\% $\pm$ 6\% & 2 \\
SMC & y & 32\% $\pm$ 5\% & 4 & 15\% $\pm$ 8\% & 2 \\ 
LMC & y & 21\% $\pm$ 5\% & 10 & 3\% $\pm$ 1\% & 1 \\
MW ext. & y & 18\% $\pm$ 7\% & 5 & \nodata & \nodata \\
MW int. & y & 11\% $\pm$ 5\% & 3 &  \nodata & \nodata \\
SMC & o & 27\% $\pm$ 8\% & 2 & 21\% $\pm$ 7\% & 2 \\
LMC & o & 16\% $\pm$ 2\% & 1 & 25\% $\pm$ 4\% & 1 \\
\tableline
\tablecomments{The fractional Be content of clusters having similar ages and metallicities
in Table \ref{earlylate} were averaged to produce the data in
this table.  Clusters with fewer than 20 objects in a spectral bin were not used in the averaging 
procedure.  The column labeled size corresponds to the sample size of clusters used to compute these averages.}
\label{agezstats}
\end{tabular}
\end{table}

\clearpage
\begin{table}
\caption{Main Sequence Lifetimes of Early B-Type Stars}
\scriptsize
\begin{tabular}{cccccc}
\tableline
Mass &  Initial Rot. Velocity & Mid-MS Lifetime & Mid-MS Lifetime & MS Lifetime & MS Lifetime \\
         &                                      &    SMC         &      Solar      &         SMC   & Solar \\
\tableline
25 M$_{sun}$ ($\sim$O9) & 0 & 3.6 Myr & 3.0 Myr & 7.2 Myr & 5.9 Myr \\
\nodata & 300 km s$^{-1}$ & 3.9 Myr & 3.7 Myr & 7.8 Myr & 7.4 Myr \\
15 M$_{sun}$ ($\sim$ B0) & 0 & 6.1 Myr & 5.1 Myr & 12.2 Myr & 10.2 Myr \\
\nodata & 300 km s$^{-1}$ & 6.8 Myr & 6.5 Myr & 13.6 Myr & 12.9 Myr \\
12 M$_{sun}$ ($\sim$ B1) & 0 & 8.3 Myr & 7.0 Myr & 16.6 Myr & 13.9 Myr \\
\nodata & 300 km s$^{-1}$ & 9.3 Myr & 8.4 Myr & 18.6 Myr & 16.8 Myr \\
9 M$_{sun}$ ($\sim$ B2) & 0 & 13.0 Myr & 11.1 Myr & 25.9 Myr & 22.1 Myr \\
\nodata & 300 km s$^{-1}$ & 14.7 Myr & 13.4 Myr & 29.3 Myr & 26.7 Myr \\
\tableline
\tablecomments{The mid-point main sequence lifetimes and main sequence lifetimes of rotating and
non-rotating stars in solar and low metallicity environments is tabulated, as calculated by 
\citet{mey00} and \citet{mae01}.  The general spectral types associated with each entry were
based upon the masses of individual spectral types given by \citet{lan82}: O9 = 23 M$_{sun}$, 
B0 = 17.5 M$_{sun}$, B1 = 13 M$_{sun}$, B2 = 7.6 M$_{sun}$, and B3 = 5.9 M$_{sun}$.}
\label{lifetimes}
\end{tabular}
\end{table}

\clearpage
\begin{table}
\caption{Spectral Types of Candidate Be Stars in Galactic Clusters}
\scriptsize
\begin{tabular}{cccccccccc}
\tableline
Cluster & B0 & B1 & B2 & B3 & B4 & B5 & B6 & B7 & B8 \\
\tableline
NGC 2186 & 0/0 & 1/1 & 0/0 & 0/0 & 2/2 & 0/0 & 1/1 & 1/1 & 0/0 \\
NGC 2439 & 3/8 & 1/7 & 0/14 & 0/4 & 0/10 & 0/18 & 0/22 & 1/11 & 0/19 \\
\tableline
\tablecomments{Spectral types were assigned using the calibrations of
\citet{zor97}, following the methods outlined by \citet{gre97}:
$B_{0}: M_{V} < -3.25$, $B_{1}:  -3.25<M_{V}<-2.55$, $B_{2}:
-2.55<M_{V}<-1.8$, $B_{3}: -1.8<M_{V}<-1.4$, $B_{4}: -1.4<M_{V}<-0.95$,
$B_{5}: -0.95<M_{V}<-0.6$, $B_{6}: -0.6<M_{V}<-0.25$, 
$B_{7}: -0.25<M_{V}<0.025$, $B_{8}: 0.025<M_{V}<0.285$.  For the purposes of this 
Table, we have increased the observed m$_{V}$ for our NGC 2186 data by 1.0 magnitudes 
to correct for the systematic errors known to reside in these data (see Sections 2 and 3.3.1 
for further details regarding these errors).} 
\label{galacticspectypes}
\end{tabular}
\end{table}

\end{document}